\documentclass[fleqn,usenatbib]{mnras}

\usepackage{newtxtext,newtxmath}

\usepackage[T1]{fontenc}

\DeclareRobustCommand{\VAN}[3]{#2}
\let\VANthebibliography\thebibliography
\def\thebibliography{\DeclareRobustCommand{\VAN}[3]{##3}\VANthebibliography}

\usepackage{graphicx}	
\usepackage{amsmath}	
\usepackage{longtable}

\usepackage{soul}

\title[Individual subpulses of PSR B1916+14]
{Individual subpulses of PSR B1916+14 and their polarization properties}

\author[T. Wang et al.]{
Tao Wang,$^{1}$
C. Wang,$^{1}$ 
J.~L. Han,$^{1,2,3}$\thanks{E-mail: hjl@nao.cas.cn}
N. N. Cai,$^{1}$
W. C. Jing,$^{1}$
Yi Yan,$^{1}$
and P.~F. Wang$^{1}$
\\
$^{1}$National Astronomical Observatories, Chinese Academy of Sciences, Jia20 Datun Road, Beijing 100012, China\\
$^{2}$School of Astronomy and Space Science, University of Chinese Academy of Sciences, 19A Yuquan Road, Beijing 100049, China\\
$^{3}$The FAST key laboratory, Chinese Academy of Sciences, Jia 20 DaTun Road, Beijing 100012, China
}

\date{Accepted XXX. Received YYY; in original form ZZZ}

\pubyear{2015}

\begin{document}
\label{firstpage}
\pagerange{\pageref{firstpage}--\pageref{lastpage}}
\maketitle

\begin{abstract}
Individual subpulses of pulsars are regarded as the basic emission components, providing invaluable information to understand the radio emission process in the pulsar magnetosphere. Nevertheless, subpulses are overlapped with each other along the rotation phase for most pulsars, making it difficult to study the statistical properties of subpulses. Among the pulsars observed by the Five-hundred-meter Aperture Spherical radio Telescope, PSR B1916+14 has a large number of isolated well-resolved subpulses in the high time resolution observations, having a typical width of 0.15 ms and a high linear polarization. We find that the number distribution of subpulses contributes dominantly to the mean profile. According to the emission geometry, these emission units come from a region roughly 155 km above the polar cap in the pulsar magnetosphere, and the length scale of basic emission units is approximately 120~m. The deviations of polarization position angles for these single subpulses from the standard S-shaped curve are closely related to their fractional linear and circular polarization, and the large deviations tend to come from drifting subpulses.
\end{abstract}



\section{Introduction}

The pulsar radio emission is polarized. The mean pulse profile delineates the emission window in the pulsar magnetosphere \citep{lm+1988}, and the polarization position angle (hereafter PA) at each rotation phase corresponds to the plane of curved magnetic field lines according to the Rotation Vector Model \citep[hereafter RVM,][]{rc+1969, komesaroff+1970}. An individual pulse consists of many subpulses emerging at different phases with various strengths. \citet{scr+1984,scw+1984} studied the statistical properties of single pulses and found the orthogonal polarization modes. Mode-separated pulse profiles are often highly polarized \citep{ms+2000}, which implies a depolarization caused by overlapped subpulses with different polarization modes.
Distinguishable subpulses (or single subpulses), such as micro-pulses and giant pulses, have been detected from a few pulsars, and these reflect the characters of the basic emission units coming from charged particle bunches. Both micro-pulses and giant pulses have a narrow width smaller than hundreds of microseconds, or even nanoseconds \citep{hb+1978,hkw+2003,spb+2004,jpk+2010,mar+2015,kp+2018}, and are usually highly polarized.  

The accurate polarization measurements of narrow single subpulses give first-hand information on emission processes, as demonstrated by the detected dwarf pulses from PSR B2111+46 in nulling mode \citep{cyh23}, providing the following observation conditions: 1) a high time resolution enabled by a good digital backend; 2) a good quality of polarization calibration; 3) a high signal-to-noise-ratio of pulse detection; 4) a pulsar with a substantial number of well-resolved subpulses. The first three conditions can be satisfied by a large sensitive telescope, but the last one is hard to satisfy because subpulses are often overlapped for most pulsars.

The Five-hundred-metre Aperture Spherical radio Telescope \citep[FAST,][]{NRD+2006} is currently the most sensitive radio telescope for pulsar observations. It has been used for the FAST Galactic Plane Pulsar Snapshot (GPPS) survey  \citep{HWW+2021}, the most efficient pulsar-search survey on the galactic plane currently in the world. Up to now, the survey has detected 637 new pulsars. During test observations for the survey, a large number of known pulsars have been detected. We examined high-time-resolution profiles of individual pulses and found that PSR B1916+14 is the best to have resolved single subpulses. It is a bright pulsar discovered by \citet{ht+1974}, and has a period of $\rm{P_0} = 1.181$s and $\rm{DM } =27.202 \; pc\; cm^{-3}$ \citep{hlk+2004} and ${\rm RM=-41.7\pm3.5~rad~m^{-2}}$\citep{hmv+2018}. 
The mean pulse profile is very narrow with a full width of 28~ms at the half maximum at 1.3~GHz, occupying only 2.4 percent of its period \citep{hfs+2004}. This narrow profile has three predominant components \citep{bcw+1991, rankin+1993, wcl+1999}. The emission geometry has been determined by  \citet{omr+2019} by fitting the observed PA with a standard S-shape PA curve according to the RVM. The inclination angle of the magnetic field axis to the rotation axis is $\alpha =79 ^\circ$ and the impact angle, the closet approach between the line of sight and the magnetic axis, 
is $\beta =1.2^\circ$ \citep{rankin+1993,omr+2019}. Some subpulses are found to drift, with a drifting rate of $P_3 = 59$ periods and the phase-shift in a period being $P_2= -3.9^\circ$ \citep{swz+2023}. 

In this paper, we present the sensitive polarization observations for individual subpulses of PSR B1916+14 observed by the FAST during the GPPS survey. In Section \ref{sec:data}, the FAST observations and processing procedures are briefly introduced. In Section \ref{sec:result}, individual subpulses and their polarization properties are statistically studied. The related physics of emission and propagation processes are discussed, and conclusions are given in Section~\ref{sec:discussion}.

\begin{table}
    \caption[]{FAST observation sessions for PSR B1916+14. The offset is the pulsar position from a beam center.}
    \label{tab:obs}
    \centering
    \begin{tabular}{ccccc}
    \hline
    Obs. Date     &  FAST Beam   &  Offset &  Obs. Time     &     Periods \\ 
    {\rm (YY-MM-DD)} &   name   &    (arcmin)     &  {\rm (minute)}      &      Number            \\
    \hline
    2020-05-28     &  M11  & 1.1  & 5                    &     261 \\
    2023-03-08     &  M01  & 0.0  & 15                    &     771 \\
    \hline
    \end{tabular}
\end{table}

\begin{figure}
\centering
\includegraphics[width=0.45\textwidth]{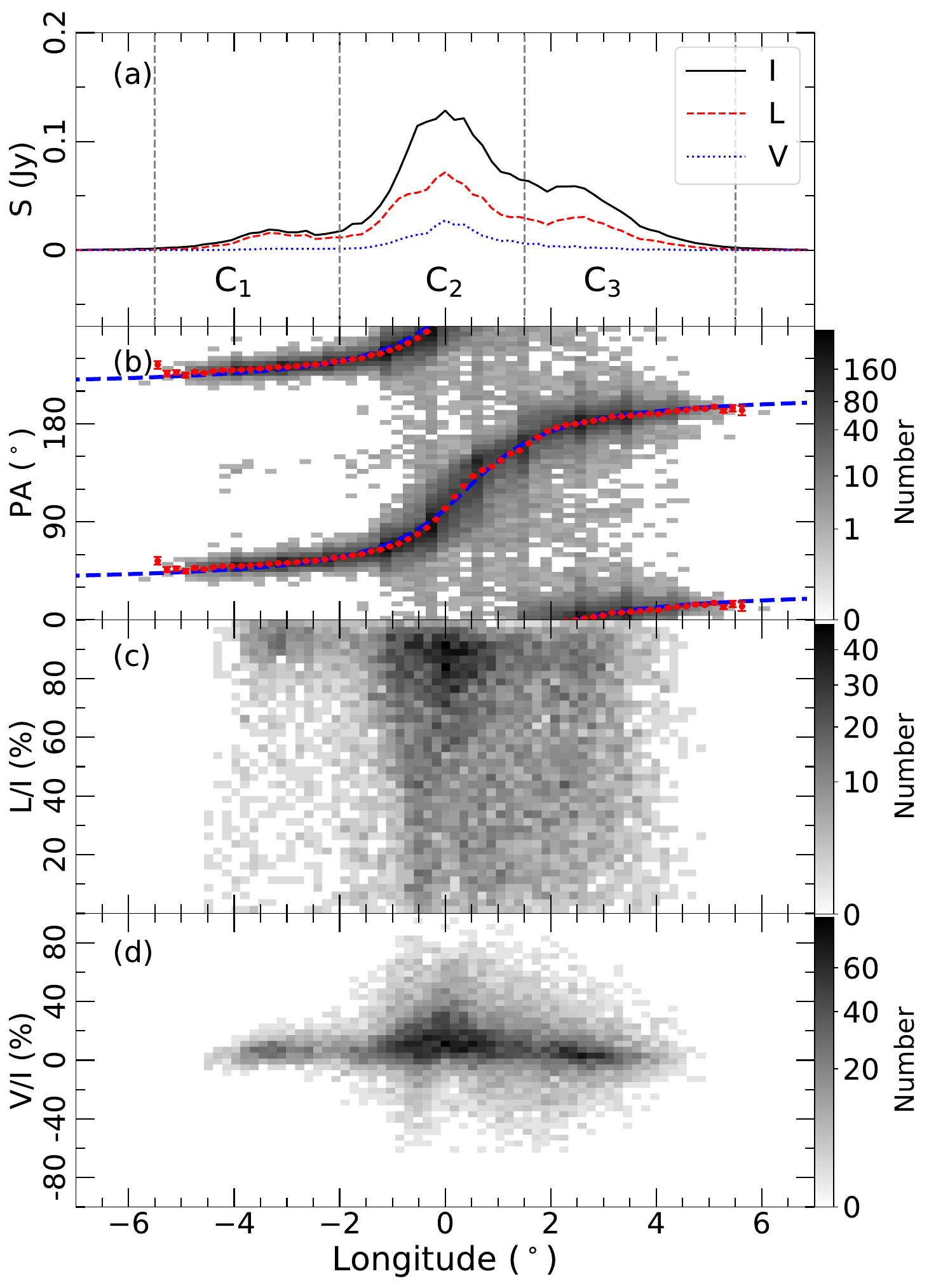}
\caption{The mean polarization profiles and the data distribution of polarized emission in the pulse window are obtained from the two observations on 2020-05-28 and 2023-03-08. In panel (a), the mean profile for the total intensity $I$, linear intensity $L$, and circular polarization $V$ are presented. The longitude ranges for the 3 mean profile components ($C_1$, $C_2$, and $C_3$) are marked. In panel (b) the data distribution of PAs for each bin of individual pulse (2048 bin per period) is plotted only when the PA has an uncertainty less than 5$^\circ$ (calculated via $\sigma_{\Psi} = (Q^2\sigma^2_{U}+U^2\sigma^2_{Q})^{1/2} /(2L^2)$. The PA curve of the mean profile (dotted line) is plotted together with the RVM fitting (dashed line). The panels (c) and (d) show the distribution of $L/I$ and $V/I$ for individual bins with $I > 20\sigma_{\rm I}$.
}
\label{AveProfPol}
\end{figure}

\section{FAST data and processing}
\label{sec:data}

The FAST observations for PSR B1916+14 were made on May 28th, 2020 (hereafter 2020-05-28) during the regular survey observations of the GPPS survey \citep{HWW+2021} by using the 19-beam L-band receivers \citep{jth+2020}. The pulsar is located in the 11th beam but $1.1'$ away from the center, see Table~\ref{tab:obs}. To verify the results on single subpulses, one more  FAST session was carried out for 15 minutes on March 8th, 2023 (hereafter 2023-03-08) by using the central beam of the 19-beam receiver. The FAST has a full gain of $\mathit{G} = 16$~K/Jy when it points to any object with a zenith angle smaller than 28.5$^\circ$. The receivers cover the frequency range from 1.0 to 1.5~GHz, and the data from all 19 beam receivers are recorded for four polarization channels of $XX$, $YY$, $Re(X^\ast Y)$ and $Im(XY^\ast)$ with a time resolution of 49.152 $\upmu$s. In case there is any known pulsar in the targeted region for a GPPS survey observation, data are recorded for two extra minutes but with the injected calibration signals of 1~K On-Off noise with a period of 2~s, which are used to calibrate the band-pass and polarization performance of the receiver.

The raw data are saved in a search mode with 2048 channels over the observation band. Using the ephemeris from the Australia Telescope National Facility Pulsar Catalogue \citep{mht+2005}, the data are de-dispersed and folded by using the open source package {\sc dspsr} \citep{dspsr}. 
To uncover the details of individual pulses in the 261/771 periods of the observations on 2020-05-08/2023-03-08, we fold data with the original sampling time of 49.152 $\upmu$s (24030 bins). To get the mean profile, observational data are also folded with 2048 bins per period, as shown in Figure~\ref{AveProfPol}. Careful data analysis processes were accomplished by the pulsar toolkit, {\sc psrchive} \citep{vdo+2012}. For example, by using the command {\sc psrzap}, the RFIs in folded data are cleaned in the two-dimensional frequency-time images interactively and the relative channels are weighted to zero. With {\sc pac}, the polarization leakage is corrected and the frequency bands are calibrated by using the solutions derived from the 2-minute calibration data. We get the observed rotation measurements from two  FAST observation sessions via {\sc rmfit}, $RM = -36.15\pm0.14~{\rm rad~m^{-2}}$ on 2020-05-08 and $RM = -34.56\pm0.10~{\rm rad~m^{-2}}$ on 2023-03-08. After the RM contribution from the earth ionosphere is estimated via {\sc IonFR} \citep{IonFR+2013}, the true $RM$ on 2020-05-08 and 2023-03-08 are $-36.57\pm0.17$ and $-37.04\pm0.21~{\rm rad~m^{-2}}$ respectively, consistent with but much improved from the previous value \citep{hmv+2018}.
 
\begin{figure*}
\centering
\includegraphics[width=0.9\textwidth]{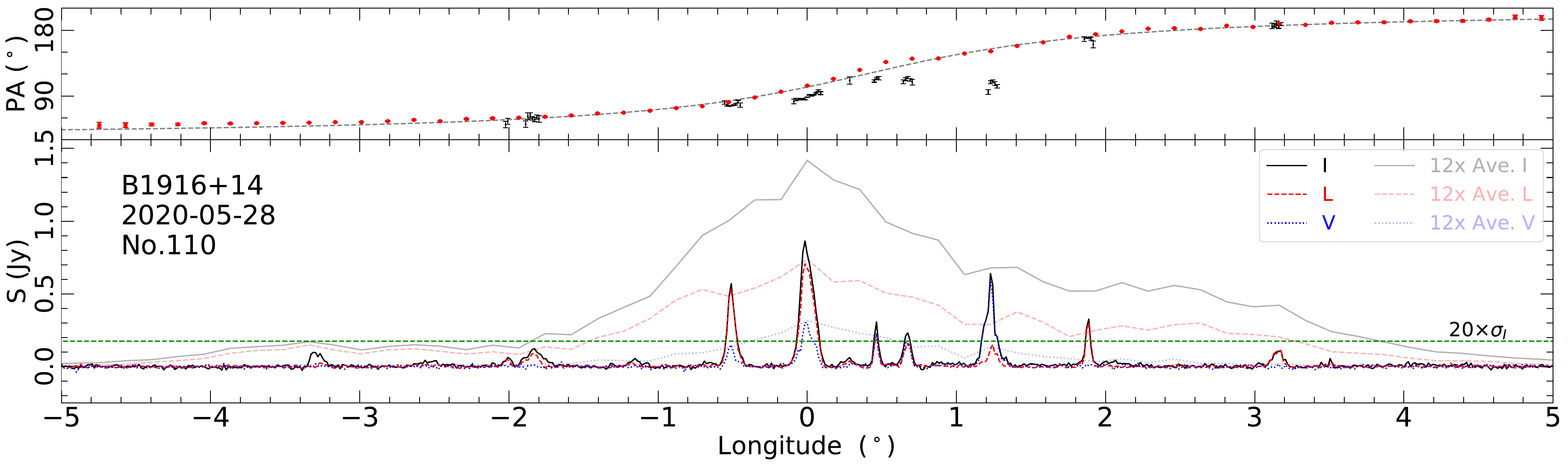}
\includegraphics[width=0.9\textwidth]{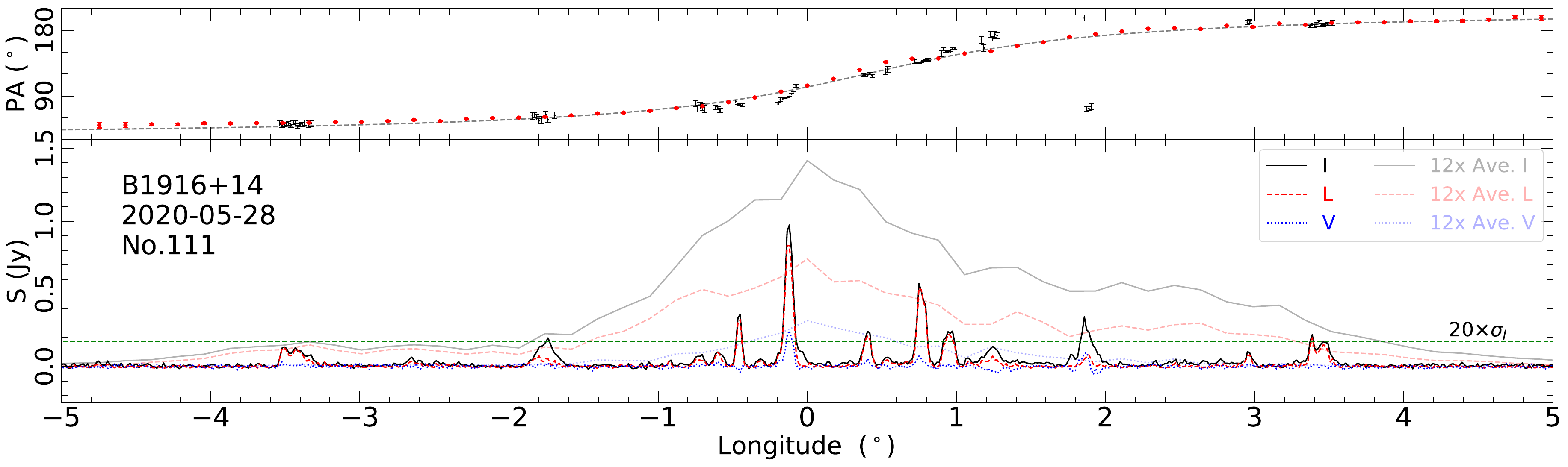}
\includegraphics[width=0.3\textwidth]{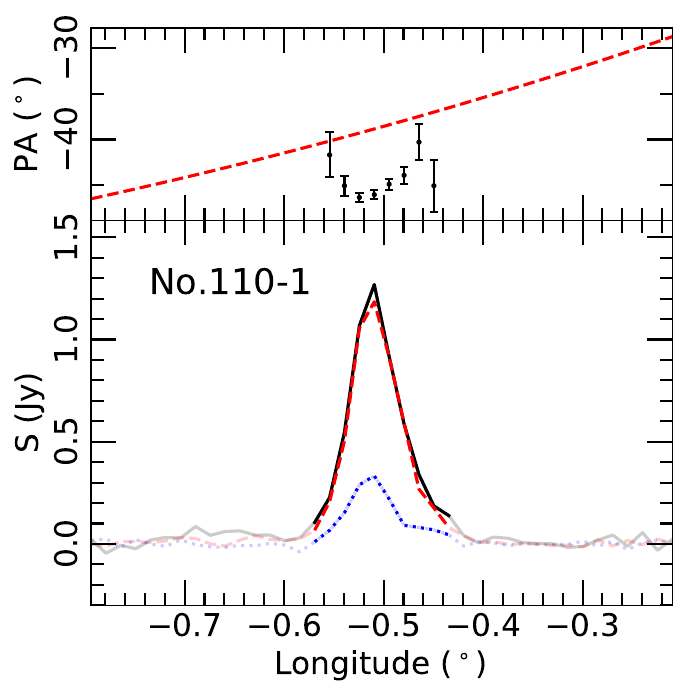}
\includegraphics[width=0.3\textwidth]{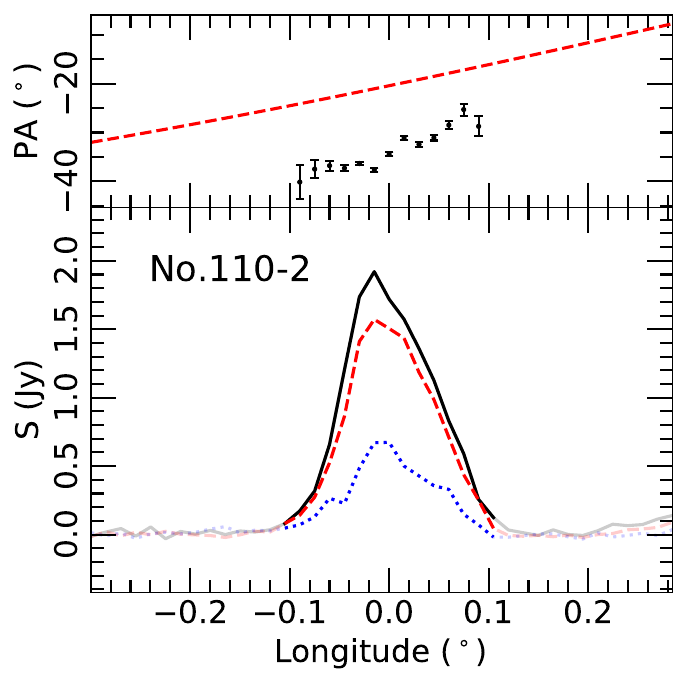}
\includegraphics[width=0.3\textwidth]{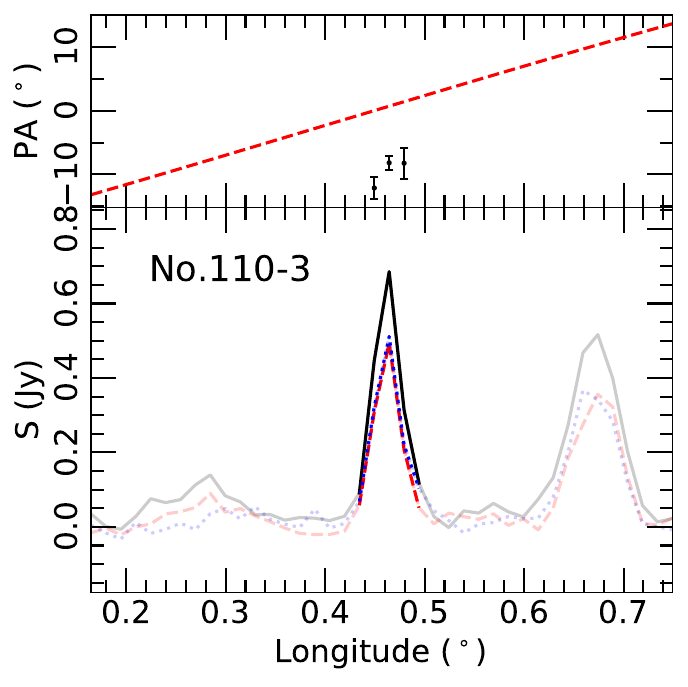}
\includegraphics[width=0.3\textwidth]{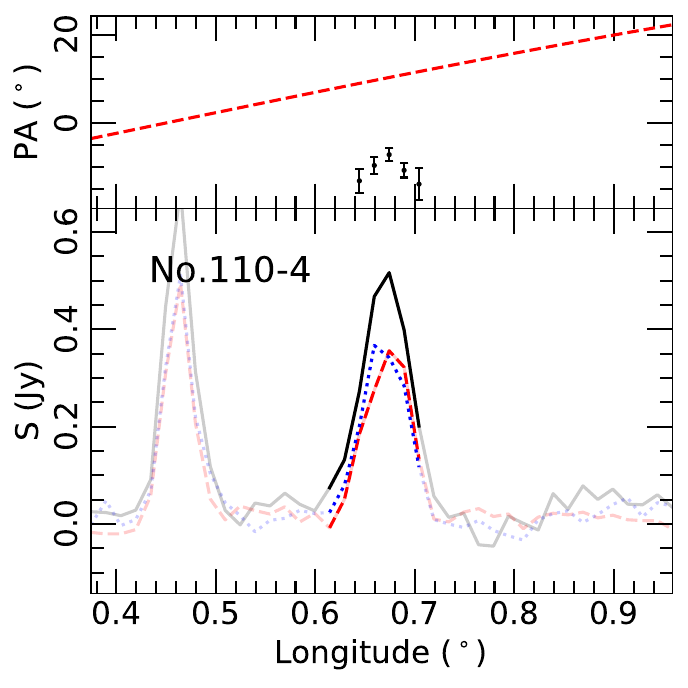}
\includegraphics[width=0.3\textwidth]{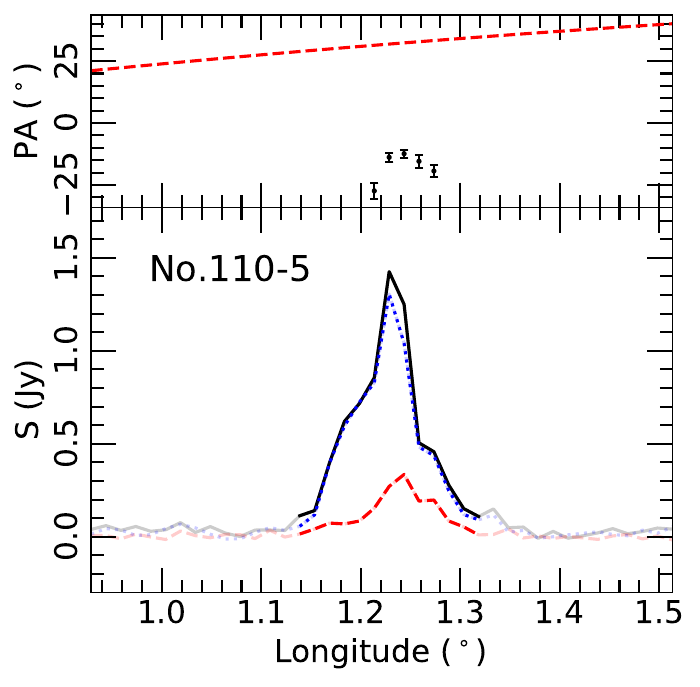}
\includegraphics[width=0.3\textwidth]{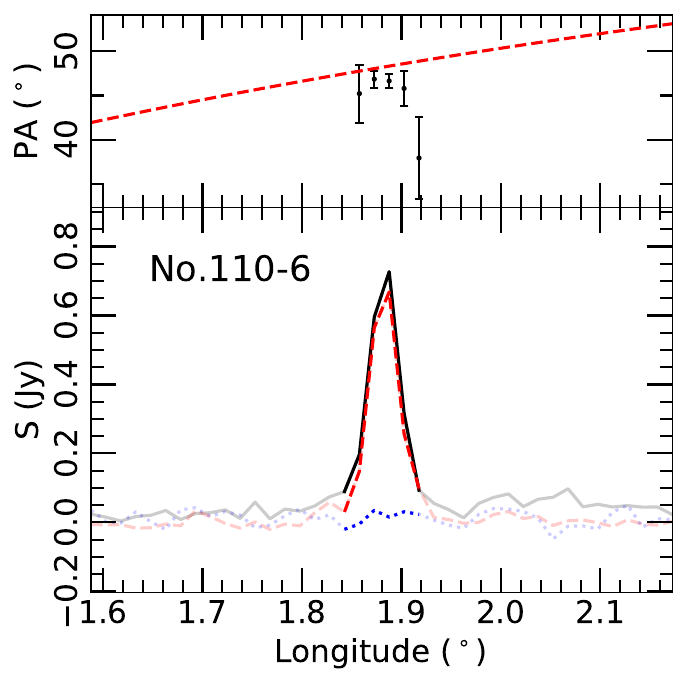}
\caption{The polarization profiles of two example individual pulses (the No.110 and 111 periods observed on 2020-05-28) with a time resolution of about 49 $\upmu$s from the super-sensitive FAST observations, with well-separated and resolved ``subpulses'' as exampled in the 6 lower panels. The total power $I$, linear polarization intensity $L$, and circular polarization intensity $V$ are shown and compared with the enlarged version (x12) of the mean polarization profiles plotted in a fainter version. The dotted line is plotted for $20\sigma_{\rm I}$ that is obtained from off-pulse data. The PA values of the individual pulses are shown in the upper subpanels if the uncertainty is less than $5^\circ$. The PA-fitted S-shape curve is plotted for comparison. The polarization profiles for these strong individual subpulses are highlighted for $I> 3\sigma_{\rm I}$. All data for the high-time-resolution profiles for individual pulses are presented in the Appendix.
}
\label{MPs}
\end{figure*}

\begin{figure*}
\centering
\includegraphics[width=0.28\textwidth]{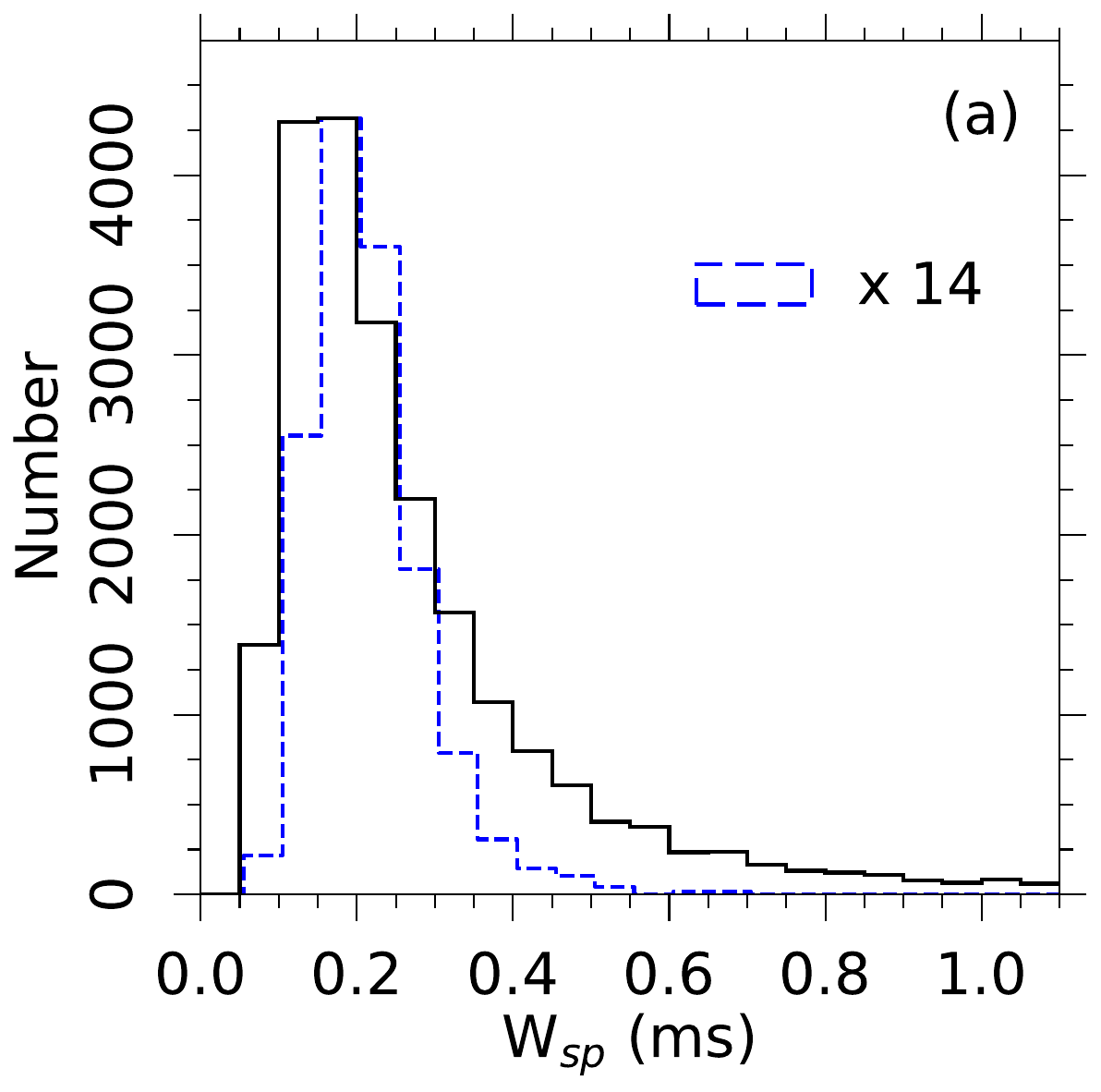}
\includegraphics[width=0.28\textwidth]{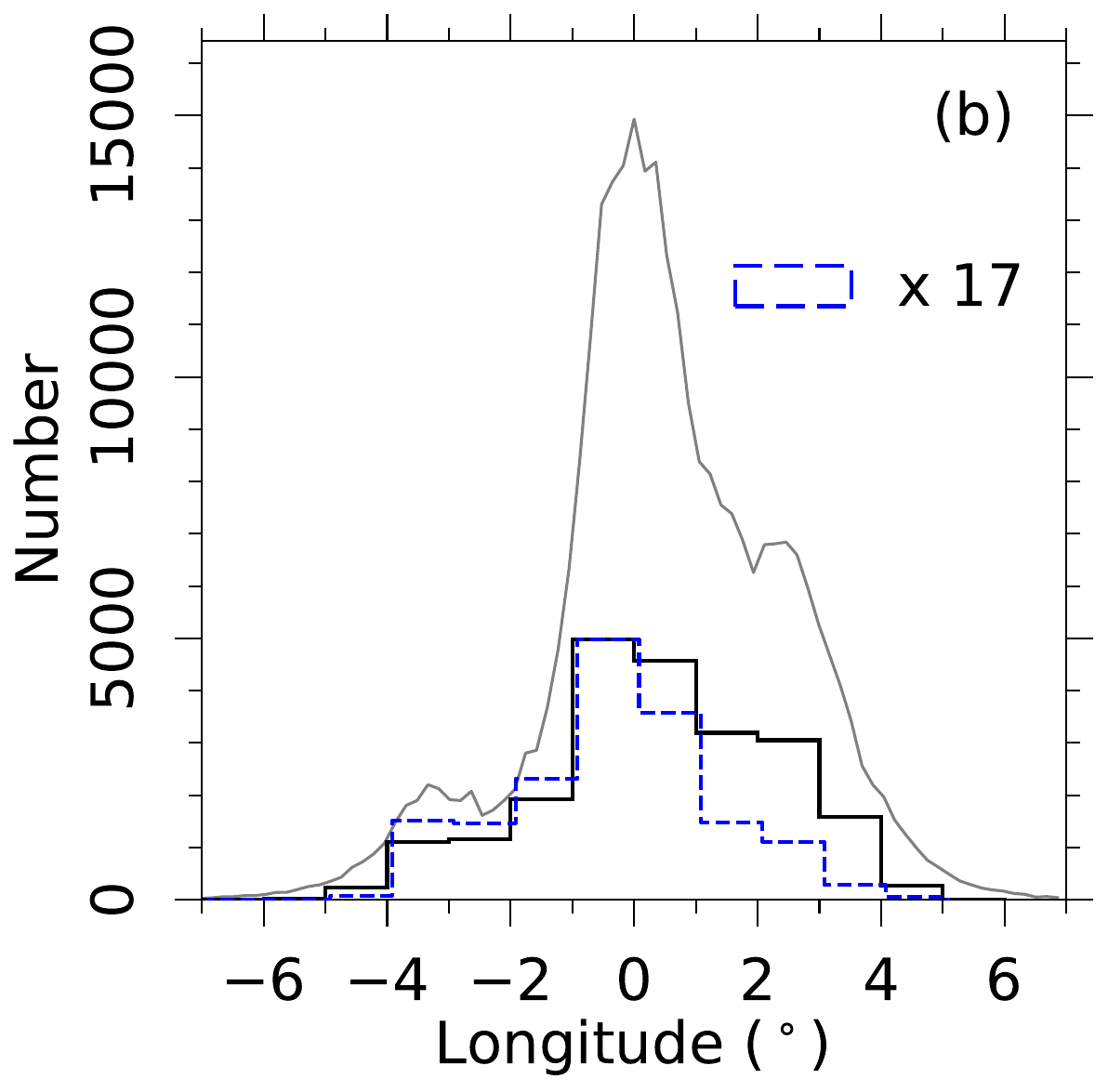}
\includegraphics[width=0.28\textwidth]{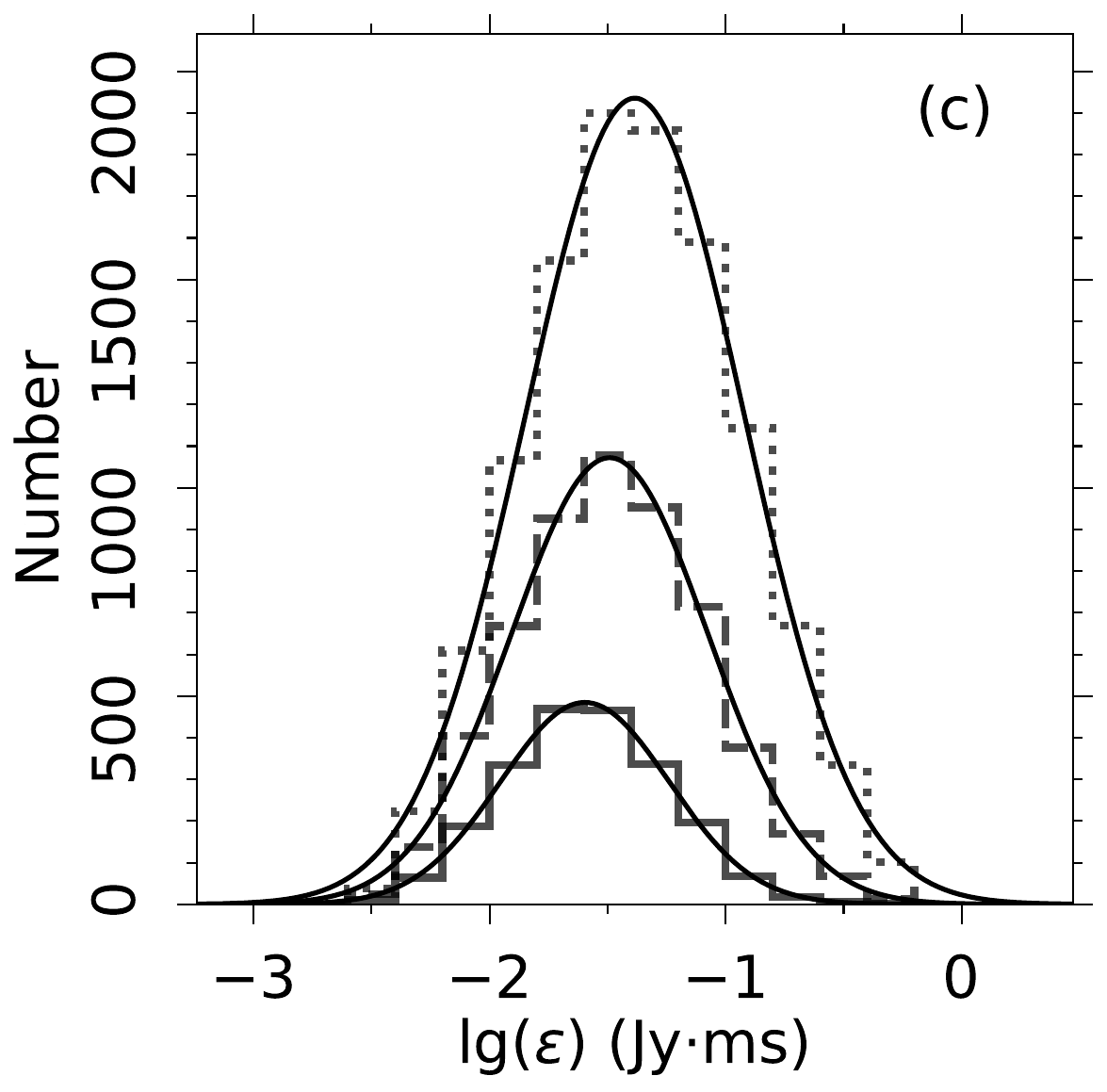}
\caption{Statistical properties of subpulses obtained by the FAST. (a) The histogram of subpulse width ${\rm W_{sp}}$ for all (solid boxes) and recognized single (dashed boxes) subpules. (b) The phase distribution for all (solid boxes) and single  (dashed boxes) subpulses, compared with the mean pulse profile (solid grey line) in  Fig.~\ref{AveProfPol}.  (c) The histograms of subpulse energy. The solid, dotted, and dashed histograms represent those of $\rm C_1$, $\rm C_2$, and $\rm C_3$ components with their fitting lines, respectively.
}
\label{distr3para}
\end{figure*}

\section{FAST observation results}
\label{sec:result}

All polarization profiles, both the mean profile and individual pulses, are obtained after removing the effect of RMs and averaging over all channels by using {\sc pam}. 

\subsection{The mean polarization profile and emission geometry}

The mean polarization profile of PSR B1916+14 observed by FAST on 2020-05-28 and 2023-03-08 are shown in Figure~\ref{AveProfPol}, which are consistent with that in \citet{wcl+1999} observed by using the Arecibo telescope. Three components are predominant in the mean profile, marked as $\rm C_1$, $\rm C_2$ and $\rm C_3$ for the leading, central, and trailing components in the phase range of [-5.5$^\circ$, -2$^\circ$], [-2$^\circ$, +1.5$^\circ$], [+1.5$^\circ$, +5.5$^\circ$], respectively. The fractional linear/circular polarization, $L/I$ and $V/I$ of these three components are 73\% and 6\%, 53\% and 15\%,  and 48\% and 5\%, respectively. The $\rm C_1$ component has the strongest linear polarization, and the $\rm C_2$ component has the strongest circular polarization without sign reversal.

Because of the super-sensitivity of FAST, the polarization properties of individual pulses have been well-measured. The PA values in the sub-panel (b) of Figure~\ref{AveProfPol} are around the mean S-shape curve, consistent very well with the RVM model (the dashed line) which is plotted with parameters of $\alpha$=79$^\circ$ and $\beta$=1.2$^\circ$ given by \citet{rankin+1993} and \citet{omr+2019}. The fitted RVM curve can be taken as a standard PA curve. A small number of bins around the longitude of $-4^\circ$ in the panel (b) of Figure~\ref{AveProfPol} show the orthogonal mode.  

The emission beam of PSR B1916+14 complies with the cone+cone model for three reasons: (1) the width evolution of the central profile component with frequency is consistent with that of the cone (see Fig.A18 of \citealt{rwv+2023}), indicating that the central profile component comes from conal emission; (2) the subpulse drifting (see Section~\ref{drifting}) detected from the central component is a common property of conal emission \citep{Rankin+1986,rwv+2023}; (3) no sign reversal of circular polarization is detected for the central component, indicating that the central profile id not the core emission from the beam center \citet{lm+1988}. Therefore, the three predominant components of the mean profile of PSR B1916+14 should be produced by the line of sight grazing the inner and outer cones of the emission beams, as discussed by \citet{omr+2019}.

The emission height of pulsars, $r_{\rm em}$, could be estimated by \citep{psrAstro}:
$$
\begin{cases}
    \rho_{\rm b} = 2\sin^{-1}[\sin^2\frac{W_{10}}{4} \sin\alpha \sin(\alpha+\beta)+\sin^2(\frac{\beta}{2})]^\frac{1}{2}\\
    r_{\rm em}  = 6.5\,{\rm km} F_{\rm lol} (P_0/1s) (\rho_{\rm b}/1^\circ)^2
\end{cases}
$$
Here $\rho_{\rm b}$ is the angular radius of the outer cone of the emission beam, and $W_{10}$ represents the pulse width at the level of 10\% of the peak. The coefficient $F_{\rm lol}=1+[2\tan\alpha/(3+\sqrt{9+8\tan^2\alpha})]^2$ describes the farthest point on the last open field line in the unit of the light cylinder radius. For PSR B1916+14,  $W_{10} = 7.57 ^{\circ}$, $F_{\rm lol}=1.33$ for $\alpha$=79$^\circ$. Taking these geometric parameters into the above equation, we get the size of the emission beam, $\rho_{\rm b}\simeq 3.9^{\circ}$ for the outer cone, and then the emission height $r_{\rm em}=155$~km.

\subsection{Morphological parameters of subpulses} 
\label{sec:MPs}

We obtained high time-resolution profiles of individual pulses for 261 and 771 periods of PSRB1916+14 (see Appendix~\ref{sec:SP}). As shown in Figure~\ref{MPs} for two examples, an individual pulse consists of many individual subpulses, which are well-separated and act as the isolated emission units. We checked individual pulse profiles for many bright pulsars observed by the FAST and found that PSR B1916+14 is the only pulsar having so large number of well-distinguished subpulses. Therefore this is an excellent chance to study the properties of individual subpulses radiated by the individual clusters of particles in the pulsar magnetosphere. 

For this purpose, we fit individual sub-pulses with the multi-Gaussian function and finally get 23321 Gaussian fittings to subpulses, among which there are 955 single subpulses (see Sec.~\ref{sec:pol}). The morphological parameters, such as the peak phase in the longitude, the pulse energy ($\varepsilon$), and the full-width at half maximum ${\rm w_{\rm sp}}$, are obtained and published online (see Table~\ref{tabB1} in Appendix~\ref{sec:SSP}).

The distributions of these morphological parameters for the 23321 subpulses are given in Figure~\ref{distr3para}. These subpulses have a typical width ${\rm w_{\rm sp}}$ in the range from 0.05~ms to 1.0~ms and peak at about 0.15~ms (solid line). For a typical width $w_{\rm sp}=0.15^{+0.85}_{-0.10}$ ms, the typical length scale of such basic emission units in the magnetosphere is estimated as $r_{\rm em}\sin\alpha\,2\pi w_{\rm sp}/P_0=$ $0.12^{+0.68}_{-0.08}$~km. The scattering time of this low DM pulsar is 0.03 $\upmu$s, and the dispersion smearing time of a channel at the central frequency of 1.25 GHz is about 27 $\mu$s, both are negligible.

\begin{table}
    \caption[]{The Gaussian fitting to the log-normal distribution of pulse energy for the three components: the number $n$, and the mean energy, $\bar{\varepsilon}$, for the three components.}
    \label{tab:1}
    \centering
    \begin{tabular}{cccccc}
    \hline
    Comp.    & amplitude         &    peak $\mu$       &  width $\sigma$      &  n/$\rm n_2$  &  $\bar{\varepsilon}/\bar{\varepsilon_2}$  \\ 
                   &   ${\rm a_0} $     &  (Jy$\cdot$ms)  & (Jy$\cdot$ms)   &                &     \\
    \hline
    $C_1$          & 485$\pm$4      & -1.60$\pm$0.01  & 0.36$\pm$0.01   &    0.19        &   0.50 \\
    $C_2$          & 1935$\pm$22    & -1.38$\pm$0.01  & 0.46$\pm$0.01   &    1.00        &  1.00  \\
    $C_3$          & 1072$\pm$13     & -1.49$\pm$0.01  & 0.42$\pm$0.01   &    0.49        &  0.72  \\
    \hline
    \end{tabular}
\end{table}

The distribution of these subpulses along the rotation phase (the solid) is shown in Figure~\ref{distr3para}b, roughly following the same function as the mean profile. More subpulses are detected within the $C_2$ component near the profile center than others. The histograms of the subpulses energy ($\varepsilon$) in the logarithm format with a base of 10 for three components are shown in Figure~\ref{distr3para}c, and they are fitted well by a log-normal distribution function, which suggests a linear wave growth process \citep{cjd+2004}:
\begin{equation}
f(x) = \rm{a_0} \cdot {\rm exp}(\frac{({\rm lg}(\varepsilon)-\mu)^2}{-2\sigma^2})
\label{equ:nor}
\end{equation}
Finally, the fitting parameters are shown in Table~\ref{tab:1}. The $C_2$ component has the biggest values of $\mu$ and $\sigma$, which implies that subpulses tend to get stronger in the central component than in others. The subpulse energy is only a factor of two compared to those of the other two components. However, the number of subpulses represented by ${\rm a_0}$ in Table~\ref{tab:1} dominates more effectively to the strength of this mean profile component.

\begin{figure}
\centering
\includegraphics[width=0.42\textwidth]{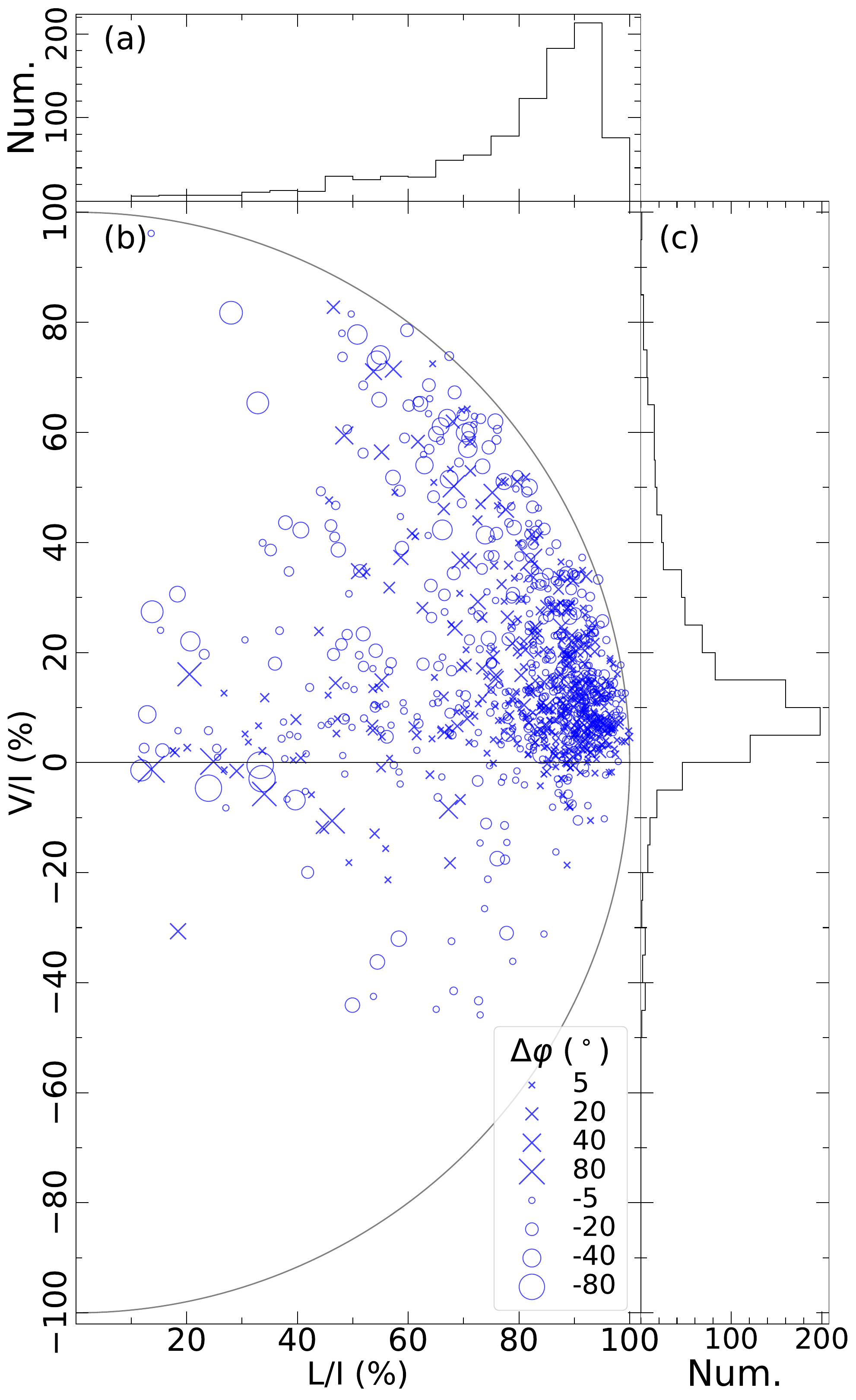}
\caption{The distribution of $L/I$ versus $V/I$ for single subpulses (subpanel {\it b}) and their number distributions (subpanels {\it a} and {\it c}). In subpanel (b), the symbol size is proportional to the PA deviation of a single subpulse from the fitted mean PA curve,  ``$\times$'' for positive offset and the circles for negative deviations. }
\label{LVplane}
\end{figure}

\begin{figure}
\centering
\includegraphics[width=0.4\textwidth]{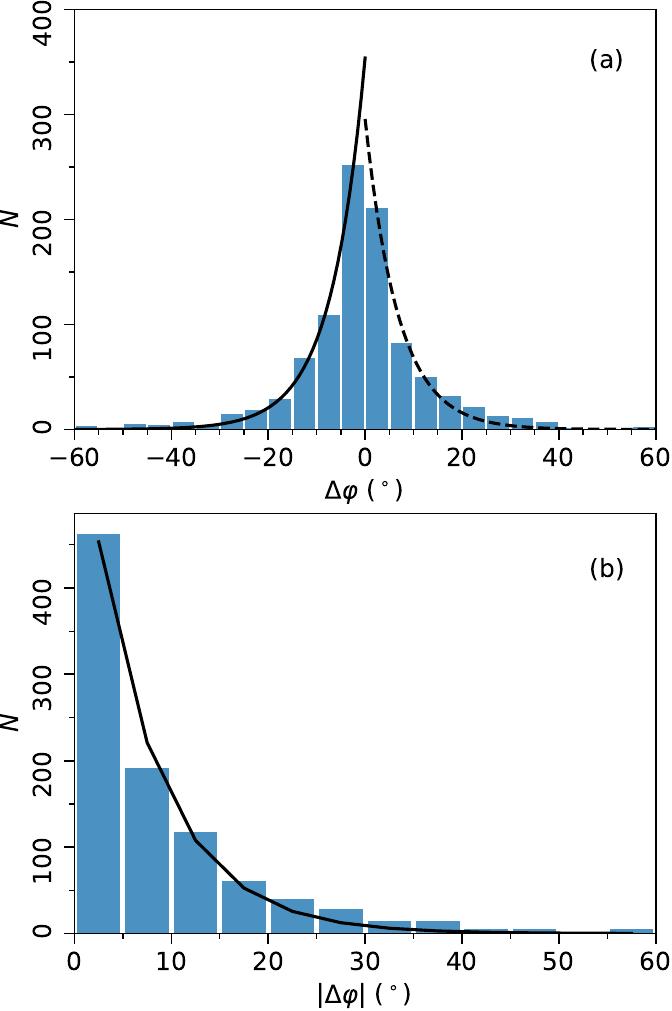}
\caption{The distribution of PA deviations from the fitted PA curve ($\Delta\varphi$ and |$\Delta\varphi$|) can be described by $ N = N_0 2^{-|\Delta\varphi|/\Delta\varphi_{\rm scale}}$. Data points of  $\Delta\varphi<0$ and  $\Delta\varphi>0$ are fitted separately in panel (a),  with $N_0=353.8\pm12.0$ and $\Delta\varphi_{\rm scale}=4.9\pm0.2$ for the negative deviations, and $N_0=295.7\pm16.8$, $\Delta\varphi_{\rm scale}=4.8\pm0.4$ for positive deviations. In panel (b), the fitting is made jointly,  and we get $N_0= 649.2\pm27.7$, $\Delta\varphi_{\rm scale}=4.8\pm0.3$.
}
\label{dPAdistr_total}
\end{figure}

\begin{figure*}
\centering
\includegraphics[width=0.8\textwidth]{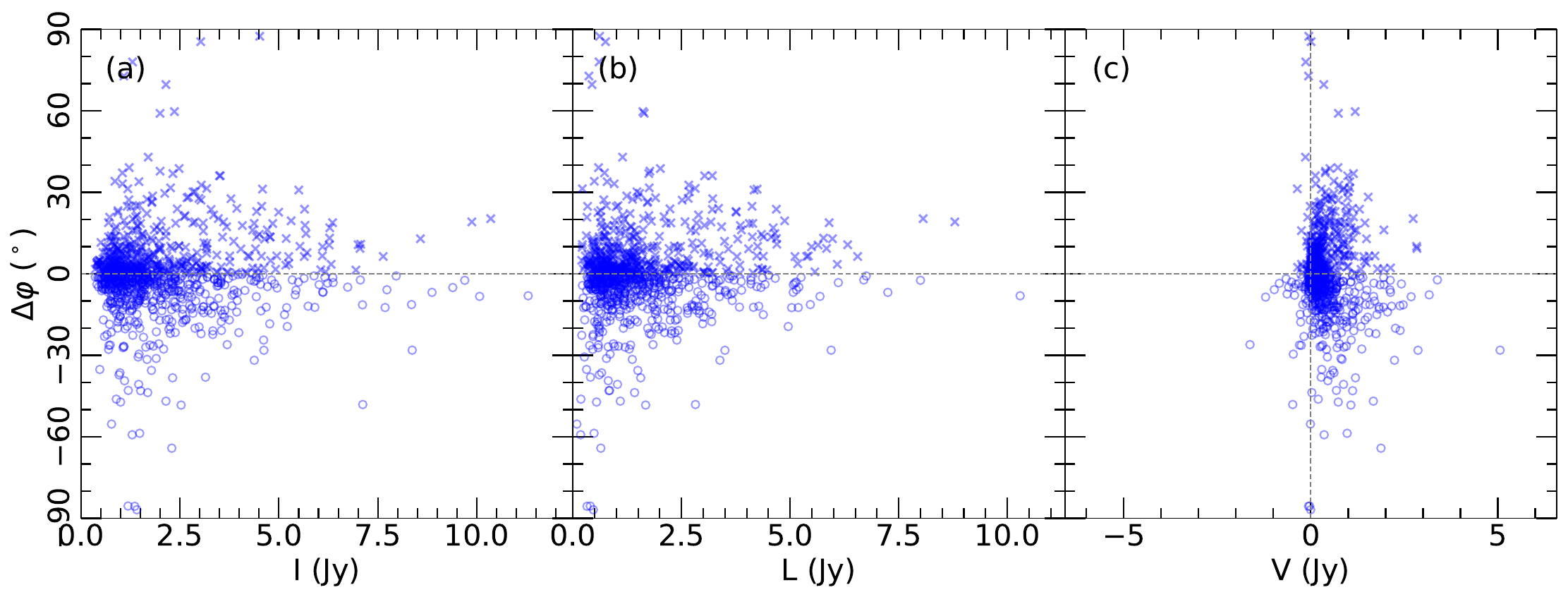}
\includegraphics[width=0.8\textwidth]{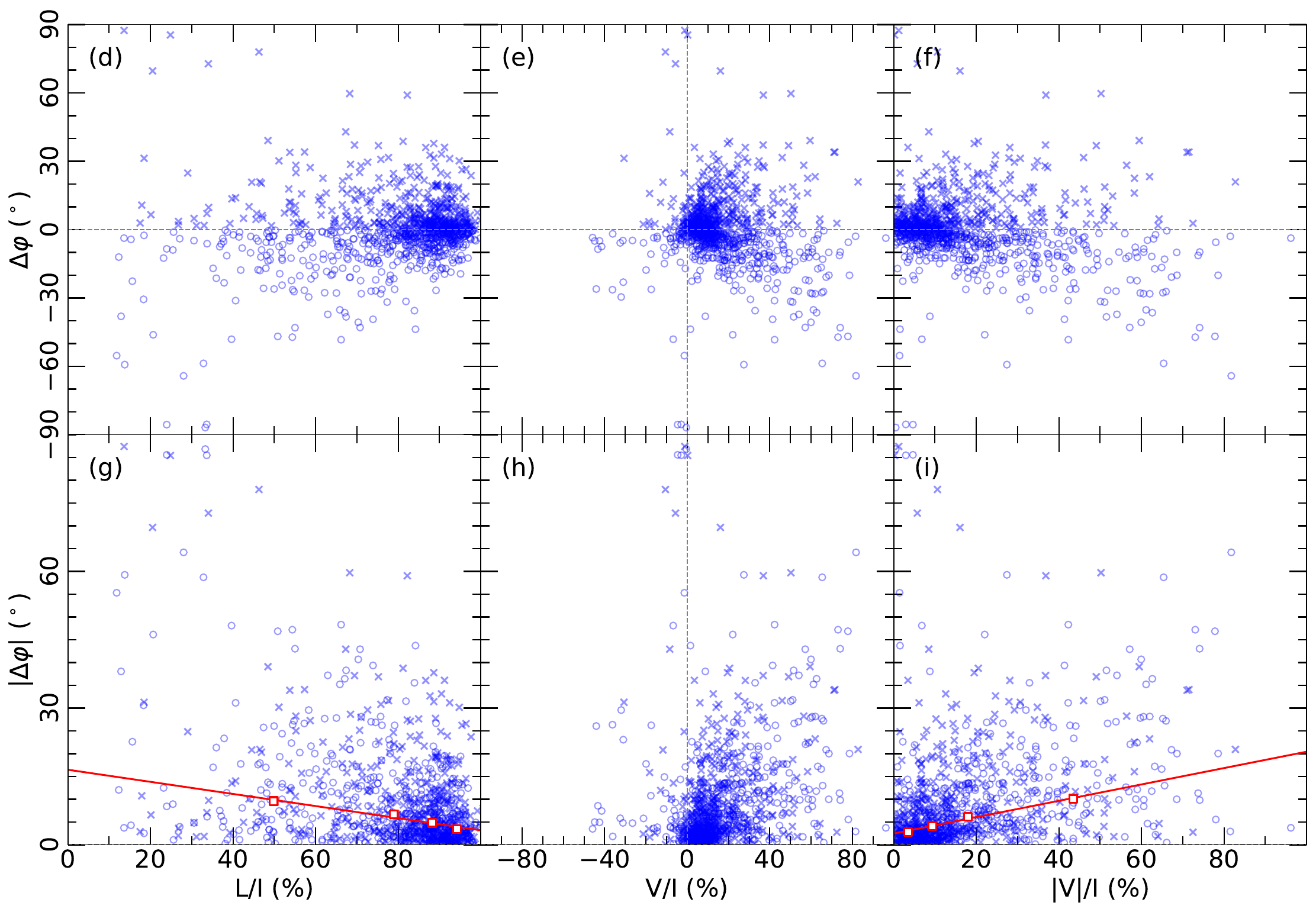}
\caption{Distributions of PA deviations ($\Delta\varphi$) against (a) $I$, (b) $L$, (c) $V$, (d) $L/I$, (e) $V/I$, and (f) $|V|/I$. The sub-panels (g), (h), and (i) show the distribution of absolute PA deviation against $L/I$, $V/I$, and $|V|/I$ respectively. The symbols ``$\times$'' and the circles represent the positive and negative PA deviations. The mean of PA deviations ($\Delta\varphi_{\rm scale}$) are obtained for 4 equal data numbers (the red boxes) for fitting them separately, which are shown in the subpanel (g) and (i).}
\label{PAvsILV}
\end{figure*}

\begin{figure}
\centering
\includegraphics[width=0.35\textwidth]{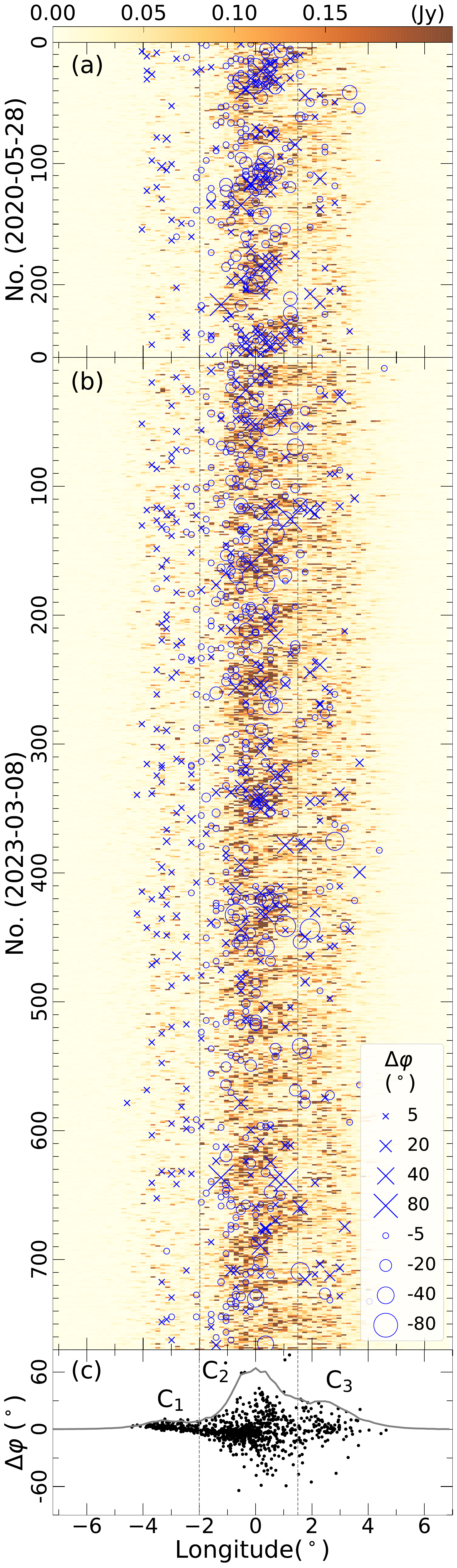}
\caption{Observational pulse sequences on 2020-05-28 (in the panel a) and 2023-03-08 (in the panel b). The single subpulses are marked with `$\times$'  ($\Delta \varphi>0$) and `$\bigcirc$' ($\Delta \varphi>0$) in both two panels for PA deviations, and symbol size is proportional to the magnitude of their |$\Delta \varphi$|. Furthermore, the $\Delta \varphi$ distribution is shown in panel (c) with the mean pulse profile in a grey line.}
\label{shifting}
\end{figure}

\subsection{Polarization properties of single subpulses} 
\label{sec:pol}

We pick up well-separated single-subpulses to avoid any confusion of overlapped subpulses for better statistics, as demonstrated in Figure~\ref{MPs} for the 110th period observed on 2020-05-28. The single-subpulses of PSR B1916+14 are selected with the following conditions: (1) peak intensity greater than 20$\sigma_{\rm I}$, so that the subpulse is strong enough to ensure a high confidence for the polarization parameter determination; (2) subpulses nearly isolated as chosen by the algorithm introduced in Appendix~\ref{sec:SSP}. Those with obvious shoulder components such as the No.~110-5 in Figure~\ref{MPs} are discarded, because the shoulder components may affect the width measurements and polarization measurements for the concerned peak. In the end, from individual pulses of  1032 periods of PSR B1916+14, we get 955 single subpulses with high-quality polarization measurements. 

The single-subpulses, as basic emission units, often have strong linear polarization (see Figure~\ref{LVplane}). More intriguing is that the large PA deviations from the standard PA curve are seen from some single-subpulses, see examples in Figure~\ref{MPs}. 
To explore the polarization properties and their interplay for these individual subpulses as basic emission units, we measure the polarization parameters for all single-subpulses, as listed in Table~\ref{tabB1} in the Appendix~\ref{sec:SSP}, including the peak flux density $I_{\rm p}$, linear polarization $L_{\rm p}$, circular polarization intensity $V_{\rm p}$ (positive for the left-hand sense) and the absolute value $|V_{\rm p}|$. For each of single-subpulses, data bins in the longitude range of ${\rm w_{sp}}$ are summed to get the integrated flux density for a subpulse on $I$,  $L$, $V$ and $|V|$. Other derived parameters, linear polarization percentage (hereafter, $L/I$), and circular polarization percentage (hereafter, $V/I$) are all derived for the integrated flux density for every single subpulse. 

As seen in Figure~\ref{LVplane}, most single subpulses are highly linearly polarized with weak circular polarization, even though a small number of them are highly circularly polarized, as indicated by the histograms shown in the subpanels (a) and (c). Their histograms peak at $L/I$=90\% and $V/I$=10\%. The PA deviations are somehow related to $L/I$ and also $V/I$. We then check observation data in Figure~\ref{PAvsILV}, see if there is any dependence of $\Delta \varphi$ on $I$, $L$, and $V$. 

The PA deviation ($\Delta\varphi$) of a single subpulse from the standard S-shape curve is the weighted average for values in these bins, calculated by 
\begin{equation}
    {\Delta\varphi} = \sum_{i=1}^n (\varphi^i-\varphi^i_0)\sigma^{-2}_{\varphi^i} / \sum_{i=1}^n \sigma^{-2}_{\varphi^i}
\label{equ:1}
\end{equation}
in which $\varphi^i$ represents the PA value of the $i$th bin, and $\sigma_{\varphi^i}$ is the uncertainty of $\varphi^i$ and only those bins with an error less than 5$^\circ$ are considered. Such a weighted average naturally rests on the values in better-measured bins inside a subpulse.

We analyzed the $\Delta \varphi$ distribution of all the single-subpulses and found that it fits an exponential distribution quite well as
\begin{equation}
    N = N_0 2^{-|\Delta\varphi|/\Delta\varphi_{\rm scale}}
\label{equ:dPAdistr}
\end{equation}
for $\Delta\varphi>0$, $\Delta\varphi<0$ or $|\Delta\varphi|$ as shown in Figure~\ref{dPAdistr_total}. Here the scale of the exponential distribution $\Delta\varphi_{\rm scale}$ reflects the dispersion degree of PA deviations. The single-subpulses with $\Delta\varphi<0$ and $>0$ have $\Delta\varphi_{\rm scale}=4.9^\circ\pm0.2^\circ$ and $4.8^\circ\pm0.4^\circ$ respectively. For all the single-subpulses, $\Delta\varphi_{\rm scale}=4.8^\circ\pm0.3^\circ$.

The PA deviations spread over a wide range in $\pm90$, and show non-dependence on $I$ and $V$ (see Figure~\ref{PAvsILV}a and c), but are bigger for these single subpulses with weaker linearly polarized intensity (see Figure~\ref{PAvsILV}b). We further check the dependence of $\Delta \varphi$ on $L/I$ and $V/I$. The PA deviations become more dispersed as $L/I$ becomes smaller (see Figure~\ref{PAvsILV}d), in a similar style for positive and negative $\Delta \varphi$. But just in contrast, the more dispersed $\Delta \varphi$ data are found for stronger circular polarization (see Figure~\ref{PAvsILV}e and f). We find that each of the $\Delta \varphi$ distributions is symmetrical with $\Delta \varphi=0^\circ$, they are folded along $\Delta \varphi=0^\circ$ for simplicity (see Figure~\ref{PAvsILV}g to i). Quantitatively, we divide all the points into 4 equal portions in order of $L/I$ and/or |$V/I$|, and get their $\Delta \varphi_{\rm scale}$ (the red boxes in Figure~\ref{PAvsILV}g and i). The scales of PA deviation distribution are linearly related to fractional linear polarization, $\Delta \varphi_{\rm scale}=(-13.3\pm2.0)\times L/I+(16.5\pm1.5)$, and to fractional circular polarization $\Delta \varphi_{\rm scale}=(18.0\pm1.4)\times |V/I|+(2.5\pm0.3)$.

\subsection{Subpulse drifting and PA deviation}
\label{drifting}

From the FAST data of PSR B1916+14, we find that subpulse-drifting in the components $C_2$ and $C_3$, as shown in Figure~\ref{shifting}. Through the two-dimensional Fluctuation Spectra method \citep[2DFS,][]{EdwSta2002}, we obtained a drift periodicity $P_3$=$58.3 P_0$ and a longitude spacing of $P_2$=$-3.7^\circ$, consistent with the results given by \citet{swz+2023}. The large $P_3$ among the known driftings may imply a low acceleration potential drop of this pulsar according to the partially screened gap model \citep{Gil2003,bmm+2016}.

Interestingly, the single subpulses with a large $\Delta \varphi$ appear mainly in the drifting $C_2$ and $C_3$ components (see the lower panel of Figure~\ref{shifting}). The corresponding emission in these two components is relatively strong. In physics, the plasma environment in the emission region of the $C_2$ and $C_3$ components should be denser and more complicated compared to that of the $C_1$ component. We suspect the large PA deviation from the PA curve of individual subpulses may relate to the complicated plasma environment. 

\section{Discussion and Conclusion}
\label{sec:discussion}

Based on FAST observations, PSR B1916+14 is found to have a very well S-shape PA curve of the mean profile, consistent highly with the standard PA curve predicted by the RVM model with $\alpha=79^\circ$ and $\beta=1.2^\circ$ \citep{omr+2019}. Its emission beams comply with the cone+cone model and the emission height of the outer cone is estimated to be 155 km and the size of the outer cone $\rho_{\rm b} \simeq3.9^\circ$. In addition, the orthogonal modes are detected for a only very small number of subpulses. This pulsar radiates almost in a single natural wave mode (ordinary or extraordinary mode).

The basis properties of individual subpulses have been investigated by a large number of distinguishable single subpulses from the FAST super-sensitive high-time resolution observations. They have a typical width of ${\rm W_{sp}} \sim $ $0.15$ ms, corresponding to the length scale of $0.12$ km. Through investigating the distribution of phase locations and pulse energy, we found that both contribute to the mean profile, but the number of subpulses is the dominant factor. 

We find that the PA deviations of single subpulses from the standard PA curve are related to the fractional linear and circular polarization. The PA deviations are more dispersed when the fractional circular polarization becomes larger and the fractional linear polarization becomes smaller. Interestingly, the large PA deviation usually is detected from single subpulses in drifting $C_2$ and $C_3$ components.
There are four possible explanations for the PA deviations:

(1) Since the subpulses may produce at different emission heights, their PA deviations from the standard S-shape curve can be caused by the aberration and retardation effects and the possible plasma co-rotation effect \citep{bcw+1991}. Taking into account these effects, the PA dispersion around the S-shape curve may be explained. However, it fails to relate the PA deviation with the circular polarization. 

(2) Incoherent mixing of orthogonal modes may also cause the PA deviation accompanied by the strong de-polarization for both linear and circular polarization \citep{mmb+2023}. This is compatible with the weakly-polarized subpulses with a large PA deviation (see Figure~\ref{LVplane}). However, it can not explain these with strong circular polarization. 

(3) 
According to \citet{wyn+2022}, once the departure of the line of sight from the radiation beam center reaches $1/\gamma$ or more where $\gamma$ is the Lorentz factor of plasma, the significant circular polarization would be created 
and thus the calculated PA deviates from the value given by RVM curve (see Equations (29), (30) and (31) in \citealt{wyn+2022}). The fractional circular polarization would be high when the departure gets large enough, 
however, the intensity of an emission unit would decrease dramatically at the same time. That is not consistent with the finding of strong single subpulses highly-circularly polarized for example No.110 in Figure~\ref{MPs}. High fractional circular polarization can achieve hardly in the usual views like \citet{wwh+2012}.

(4) Propagation effects could be a good scenario for explaining the large PA deviation and high circular polarization for the single subpulses. According to \citealt{wlh+2010}, if the impact angle $\beta$ is enough low, 1.2$^\circ$ for PSR B1916+14, a large PA deviation would be generated accompanied by high circular polarization with a single-handedness through wave mode coupling effect by which a natural wave evolves from adiabatic to non-adiabatic in the streaming plasma along its path. As discussed in Section~\ref{drifting}, strong subpulses tend to appear in the drifting $C_2$ and $C_3$ components. That indicates a circumstance with many high-density plasma bunches, which are the basic units producing single subpulses, in the magnetosphere of the $C_2$ and $C_3$ components. When a natural wave propagates through these plasma bunches, evolving from non-adiabatic to adiabatic and finally to non-adiabatic, its large PA deviation and large circular polarization are produced. On the contrary, because there is a low-density plasma situation in the magnetosphere of the weak $C_1$ component, which leads to a weak propagation effect, the single subpulses are weakly-circularly polarized and their PAs follow the standard S-shape curve. It should be noted that a sense reversal of circular polarization could happen through the quasi-tangential effect when the radiation points are close to the conical surface scanned by the magnetic axis around the spinning axis \citep{wl+2009}. Nevertheless, the propagation processes have been discussed traditionally based on the plasma background before. Comprehensive discussions are needed for the question of how radio emission passes through many high-density plasma bunches.

In short, large PA deviation of individual subpulses can be generated by four physical mechanisms. All of them could contribute. However, the propagation effect in the pulsar magnetosphere has an advantage in explaining the dependence of PA deviation on circular polarization.

\section*{Acknowledgements}

FAST is a Chinese national mega-science facility built and operated by the National Astronomical Observatories, Chinese Academy of Sciences. The authors are supported by the National Natural Science Foundation of China (NSFC, Nos. 11988101 and 11833009).
C. Wang is partially supported by NSFC No. U1731120; 
C. Wang and P.~F. Wang are also supported by NSFC No. 12133004.

\section*{DATA AVAILABILITY}

Original FAST observation data can be accessible one year after observations. All processed data as plotted in this paper can be obtained from the authors with a kind request.



\bibliographystyle{mnras}
\bibliography{scibib} 

\begin{thebibliography}{}
\makeatletter
\relax
\def\mn@urlcharsother{\let\do\@makeother \do\$\do\&\do\#\do\^\do\_\do\%\do\~}
\def\mn@doi{\begingroup\mn@urlcharsother \@ifnextchar [ {\mn@doi@}
  {\mn@doi@[]}}
\def\mn@doi@[#1]#2{\def\@tempa{#1}\ifx\@tempa\@empty \href
  {http://dx.doi.org/#2} {doi:#2}\else \href {http://dx.doi.org/#2} {#1}\fi
  \endgroup}
\def\mn@eprint#1#2{\mn@eprint@#1:#2::\@nil}
\def\mn@eprint@arXiv#1{\href {http://arxiv.org/abs/#1} {{\tt arXiv:#1}}}
\def\mn@eprint@dblp#1{\href {http://dblp.uni-trier.de/rec/bibtex/#1.xml}
  {dblp:#1}}
\def\mn@eprint@#1:#2:#3:#4\@nil{\def\@tempa {#1}\def\@tempb {#2}\def\@tempc
  {#3}\ifx \@tempc \@empty \let \@tempc \@tempb \let \@tempb \@tempa \fi \ifx
  \@tempb \@empty \def\@tempb {arXiv}\fi \@ifundefined
  {mn@eprint@\@tempb}{\@tempb:\@tempc}{\expandafter \expandafter \csname
  mn@eprint@\@tempb\endcsname \expandafter{\@tempc}}}

\bibitem[\protect\citeauthoryear{{Basu}, {Mitra}, {Melikidze}, {Maciesiak},
  {Skrzypczak}  \& {Szary}}{{Basu} et~al.}{2016}]{bmm+2016}
{Basu} R.,  {Mitra} D.,  {Melikidze} G.~I.,  {Maciesiak} K.,  {Skrzypczak} A.,
   {Szary} A.,  2016, \mn@doi [\apj] {10.3847/1538-4357/833/1/29}, \href
  {https://ui.adsabs.harvard.edu/abs/2016ApJ...833...29B} {833, 29}

\bibitem[\protect\citeauthoryear{{Blaskiewicz}, {Cordes}  \&
  {Wasserman}}{{Blaskiewicz} et~al.}{1991}]{bcw+1991}
{Blaskiewicz} M.,  {Cordes} J.~M.,   {Wasserman} I.,  1991, \mn@doi [\apj]
  {10.1086/169850}, \href
  {https://ui.adsabs.harvard.edu/abs/1991ApJ...370..643B} {370, 643}

\bibitem[\protect\citeauthoryear{{Cairns}, {Johnston}  \& {Das}}{{Cairns}
  et~al.}{2004}]{cjd+2004}
{Cairns} I.~H.,  {Johnston} S.,   {Das} P.,  2004, \mn@doi [\mnras]
  {10.1111/j.1365-2966.2004.08067.x}, \href
  {https://ui.adsabs.harvard.edu/abs/2004MNRAS.353..270C} {353, 270}

\bibitem[\protect\citeauthoryear{{Chen} et~al.,}{{Chen} et~al.}{2023}]{cyh23}
{Chen} X.,  et~al., 2023, \mn@doi [Nature Astronomy]
  {10.1038/s41550-023-02056-z}, \href
  {https://ui.adsabs.harvard.edu/abs/2023NatAs.tmp..177C} {}

\bibitem[\protect\citeauthoryear{{Edwards} \& {Stappers}}{{Edwards} \&
  {Stappers}}{2002}]{EdwSta2002}
{Edwards} R.~T.,  {Stappers} B.~W.,  2002, \mn@doi [\aap]
  {10.1051/0004-6361:20021067}, \href
  {https://ui.adsabs.harvard.edu/abs/2002A&A...393..733E} {393, 733}

\bibitem[\protect\citeauthoryear{{Gil}, {Melikidze}  \& {Geppert}}{{Gil}
  et~al.}{2003}]{Gil2003}
{Gil} J.,  {Melikidze} G.~I.,   {Geppert} U.,  2003, \mn@doi [\aap]
  {10.1051/0004-6361:20030854}, \href
  {https://ui.adsabs.harvard.edu/abs/2003A&A...407..315G} {407, 315}

\bibitem[\protect\citeauthoryear{{Han}, {Manchester}, {van Straten}  \&
  {Demorest}}{{Han} et~al.}{2018}]{hmv+2018}
{Han} J.~L.,  {Manchester} R.~N.,  {van Straten} W.,   {Demorest} P.,  2018,
  \mn@doi [\apjs] {10.3847/1538-4365/aa9c45}, \href
  {https://ui.adsabs.harvard.edu/abs/2018ApJS..234...11H} {234, 11}

\bibitem[\protect\citeauthoryear{{Han} et~al.,}{{Han} et~al.}{2021}]{HWW+2021}
{Han} J.~L.,  et~al., 2021, \mn@doi [Research in Astronomy and Astrophysics]
  {10.1088/1674-4527/21/5/107}, \href
  {https://ui.adsabs.harvard.edu/abs/2021RAA....21..107H} {21, 107}

\bibitem[\protect\citeauthoryear{{Hankins} \& {Boriakoff}}{{Hankins} \&
  {Boriakoff}}{1978}]{hb+1978}
{Hankins} T.~H.,  {Boriakoff} V.,  1978, \mn@doi [\nat] {10.1038/276045a0},
  \href {https://ui.adsabs.harvard.edu/abs/1978Natur.276...45H} {276, 45}

\bibitem[\protect\citeauthoryear{{Hankins}, {Kern}, {Weatherall}  \&
  {Eilek}}{{Hankins} et~al.}{2003}]{hkw+2003}
{Hankins} T.~H.,  {Kern} J.~S.,  {Weatherall} J.~C.,   {Eilek} J.~A.,  2003,
  \mn@doi [\nat] {10.1038/nature01477}, \href
  {https://ui.adsabs.harvard.edu/abs/2003Natur.422..141H} {422, 141}

\bibitem[\protect\citeauthoryear{{Hobbs} et~al.,}{{Hobbs}
  et~al.}{2004a}]{hfs+2004}
{Hobbs} G.,  et~al., 2004a, \mn@doi [\mnras]
  {10.1111/j.1365-2966.2004.08042.x}, \href
  {https://ui.adsabs.harvard.edu/abs/2004MNRAS.352.1439H} {352, 1439}

\bibitem[\protect\citeauthoryear{{Hobbs}, {Lyne}, {Kramer}, {Martin}  \&
  {Jordan}}{{Hobbs} et~al.}{2004b}]{hlk+2004}
{Hobbs} G.,  {Lyne} A.~G.,  {Kramer} M.,  {Martin} C.~E.,   {Jordan} C.,
  2004b, \mn@doi [\mnras] {10.1111/j.1365-2966.2004.08157.x}, \href
  {https://ui.adsabs.harvard.edu/abs/2004MNRAS.353.1311H} {353, 1311}

\bibitem[\protect\citeauthoryear{{Hulse} \& {Taylor}}{{Hulse} \&
  {Taylor}}{1974}]{ht+1974}
{Hulse} R.~A.,  {Taylor} J.~H.,  1974, \mn@doi [\apjl] {10.1086/181548}, \href
  {https://ui.adsabs.harvard.edu/abs/1974ApJ...191L..59H} {191, L59}

\bibitem[\protect\citeauthoryear{{Jessner} et~al.,}{{Jessner}
  et~al.}{2010}]{jpk+2010}
{Jessner} A.,  et~al., 2010, \mn@doi [\aap] {10.1051/0004-6361/201014806},
  \href {https://ui.adsabs.harvard.edu/abs/2010A&A...524A..60J} {524, A60}

\bibitem[\protect\citeauthoryear{{Jiang} et~al.,}{{Jiang}
  et~al.}{2020}]{jth+2020}
{Jiang} P.,  et~al., 2020, \mn@doi [Research in Astronomy and Astrophysics]
  {10.1088/1674-4527/20/5/64}, \href
  {https://ui.adsabs.harvard.edu/abs/2020RAA....20...64J} {20, 064}

\bibitem[\protect\citeauthoryear{{Kazantsev} \& {Potapov}}{{Kazantsev} \&
  {Potapov}}{2018}]{kp+2018}
{Kazantsev} A.~N.,  {Potapov} V.~A.,  2018, \mn@doi [Research in Astronomy and
  Astrophysics] {10.1088/1674-4527/18/8/97}, \href
  {https://ui.adsabs.harvard.edu/abs/2018RAA....18...97K} {18, 097}

\bibitem[\protect\citeauthoryear{{Komesaroff}}{{Komesaroff}}{1970}]{komesaroff+1970}
{Komesaroff} M.~M.,  1970, \mn@doi [\nat] {10.1038/225612a0}, \href
  {https://ui.adsabs.harvard.edu/abs/1970Natur.225..612K} {225, 612}

\bibitem[\protect\citeauthoryear{{Lyne} \& {Manchester}}{{Lyne} \&
  {Manchester}}{1988}]{lm+1988}
{Lyne} A.~G.,  {Manchester} R.~N.,  1988, \mn@doi [\mnras]
  {10.1093/mnras/234.3.477}, \href
  {https://ui.adsabs.harvard.edu/abs/1988MNRAS.234..477L} {234, 477}

\bibitem[\protect\citeauthoryear{Lyne, Graham-Smith  \& Stappers}{Lyne
  et~al.}{2022}]{psrAstro}
Lyne A.,  Graham-Smith F.,   Stappers B.,  2022, Pulsar Astronomy, 5 edn.
Cambridge Astrophysics, Cambridge University Press,
  \mn@doi{10.1017/9781108861656}

\bibitem[\protect\citeauthoryear{{Manchester}, {Hobbs}, {Teoh}  \&
  {Hobbs}}{{Manchester} et~al.}{2005}]{mht+2005}
{Manchester} R.~N.,  {Hobbs} G.~B.,  {Teoh} A.,   {Hobbs} M.,  2005, \mn@doi
  [\aj] {10.1086/428488}, \href
  {https://ui.adsabs.harvard.edu/abs/2005AJ....129.1993M} {129, 1993}

\bibitem[\protect\citeauthoryear{{McKinnon} \& {Stinebring}}{{McKinnon} \&
  {Stinebring}}{2000}]{ms+2000}
{McKinnon} M.~M.,  {Stinebring} D.~R.,  2000, \mn@doi [\apj] {10.1086/308264},
  \href {https://ui.adsabs.harvard.edu/abs/2000ApJ...529..435M} {529, 435}

\bibitem[\protect\citeauthoryear{{Mitra}, {Arjunwadkar}  \& {Rankin}}{{Mitra}
  et~al.}{2015}]{mar+2015}
{Mitra} D.,  {Arjunwadkar} M.,   {Rankin} J.~M.,  2015, \mn@doi [\apj]
  {10.1088/0004-637X/806/2/236}, \href
  {https://ui.adsabs.harvard.edu/abs/2015ApJ...806..236M} {806, 236}

\bibitem[\protect\citeauthoryear{{Mitra}, {Melikidze}  \& {Basu}}{{Mitra}
  et~al.}{2023}]{mmb+2023}
{Mitra} D.,  {Melikidze} G.~I.,   {Basu} R.,  2023, \mn@doi [\mnras]
  {10.1093/mnrasl/slad022}, \href
  {https://ui.adsabs.harvard.edu/abs/2023MNRAS.521L..34M} {521, L34}

\bibitem[\protect\citeauthoryear{{Nan}}{{Nan}}{2006}]{NRD+2006}
{Nan} R.,  2006, \mn@doi [Science in China: Physics, Mechanics and Astronomy]
  {10.1007/s11433-006-0129-9}, \href
  {https://ui.adsabs.harvard.edu/abs/2006ScChG..49..129N} {49, 129}

\bibitem[\protect\citeauthoryear{{Olszanski}, {Mitra}  \& {Rankin}}{{Olszanski}
  et~al.}{2019}]{omr+2019}
{Olszanski} T. E.~E.,  {Mitra} D.,   {Rankin} J.~M.,  2019, \mn@doi [\mnras]
  {10.1093/mnras/stz2172}, \href
  {https://ui.adsabs.harvard.edu/abs/2019MNRAS.489.1543O} {489, 1543}

\bibitem[\protect\citeauthoryear{{Radhakrishnan} \& {Cooke}}{{Radhakrishnan} \&
  {Cooke}}{1969}]{rc+1969}
{Radhakrishnan} V.,  {Cooke} D.~J.,  1969, \aplett, \href
  {https://ui.adsabs.harvard.edu/abs/1969ApL.....3..225R} {3, 225}

\bibitem[\protect\citeauthoryear{{Rankin}}{{Rankin}}{1986}]{Rankin+1986}
{Rankin} J.~M.,  1986, \mn@doi [\apj] {10.1086/163955}, \href
  {https://ui.adsabs.harvard.edu/abs/1986ApJ...301..901R} {301, 901}

\bibitem[\protect\citeauthoryear{{Rankin}}{{Rankin}}{1993}]{rankin+1993}
{Rankin} J.~M.,  1993, \mn@doi [\apjs] {10.1086/191758}, \href
  {https://ui.adsabs.harvard.edu/abs/1993ApJS...85..145R} {85, 145}

\bibitem[\protect\citeauthoryear{{Rankin}, {Wahl}, {Venkataraman}  \&
  {Olszanski}}{{Rankin} et~al.}{2023}]{rwv+2023}
{Rankin} J.,  {Wahl} H.,  {Venkataraman} A.,   {Olszanski} T.,  2023, \mn@doi
  [\mnras] {10.1093/mnras/stac3025}, \href
  {https://ui.adsabs.harvard.edu/abs/2023MNRAS.519.3872R} {519, 3872}

\bibitem[\protect\citeauthoryear{{Soglasnov}, {Popov}, {Bartel}, {Cannon},
  {Novikov}, {Kondratiev}  \& {Altunin}}{{Soglasnov} et~al.}{2004}]{spb+2004}
{Soglasnov} V.~A.,  {Popov} M.~V.,  {Bartel} N.,  {Cannon} W.,  {Novikov}
  A.~Y.,  {Kondratiev} V.~I.,   {Altunin} V.~I.,  2004, \mn@doi [\apj]
  {10.1086/424908}, \href
  {https://ui.adsabs.harvard.edu/abs/2004ApJ...616..439S} {616, 439}

\bibitem[\protect\citeauthoryear{{Song} et~al.,}{{Song}
  et~al.}{2023}]{swz+2023}
{Song} X.,  et~al., 2023, \mn@doi [\mnras] {10.1093/mnras/stad135}, \href
  {https://ui.adsabs.harvard.edu/abs/2023MNRAS.520.4562S} {520, 4562}

\bibitem[\protect\citeauthoryear{{Sotomayor-Beltran}
  et~al.,}{{Sotomayor-Beltran} et~al.}{2013}]{IonFR+2013}
{Sotomayor-Beltran} C.,  et~al., 2013, {ionFR: Ionospheric Faraday rotation},
  Astrophysics Source Code Library, record ascl:1303.022 (\mn@eprint {ascl}
  {1303.022})

\bibitem[\protect\citeauthoryear{{Stinebring}, {Cordes}, {Rankin}, {Weisberg}
  \& {Boriakoff}}{{Stinebring} et~al.}{1984a}]{scr+1984}
{Stinebring} D.~R.,  {Cordes} J.~M.,  {Rankin} J.~M.,  {Weisberg} J.~M.,
  {Boriakoff} V.,  1984a, \mn@doi [\apjs] {10.1086/190954}, \href
  {https://ui.adsabs.harvard.edu/abs/1984ApJS...55..247S} {55, 247}

\bibitem[\protect\citeauthoryear{{Stinebring}, {Cordes}, {Weisberg}, {Rankin}
  \& {Boriakoff}}{{Stinebring} et~al.}{1984b}]{scw+1984}
{Stinebring} D.~R.,  {Cordes} J.~M.,  {Weisberg} J.~M.,  {Rankin} J.~M.,
  {Boriakoff} V.,  1984b, \mn@doi [\apjs] {10.1086/190955}, \href
  {https://ui.adsabs.harvard.edu/abs/1984ApJS...55..279S} {55, 279}

\bibitem[\protect\citeauthoryear{{Wang} \& {Lai}}{{Wang} \&
  {Lai}}{2009}]{wl+2009}
{Wang} C.,  {Lai} D.,  2009, \mn@doi [\mnras]
  {10.1111/j.1365-2966.2009.14895.x}, \href
  {https://ui.adsabs.harvard.edu/abs/2009MNRAS.398..515W} {398, 515}

\bibitem[\protect\citeauthoryear{{Wang}, {Lai}  \& {Han}}{{Wang}
  et~al.}{2010}]{wlh+2010}
{Wang} C.,  {Lai} D.,   {Han} J.,  2010, \mn@doi [\mnras]
  {10.1111/j.1365-2966.2009.16074.x}, \href
  {https://ui.adsabs.harvard.edu/abs/2010MNRAS.403..569W} {403, 569}

\bibitem[\protect\citeauthoryear{{Wang}, {Wang}  \& {Han}}{{Wang}
  et~al.}{2012}]{wwh+2012}
{Wang} P.~F.,  {Wang} C.,   {Han} J.~L.,  2012, \mn@doi [\mnras]
  {10.1111/j.1365-2966.2012.21053.x}, \href
  {https://ui.adsabs.harvard.edu/abs/2012MNRAS.423.2464W} {423, 2464}

\bibitem[\protect\citeauthoryear{{Wang}, {Yang}, {Niu}, {Xu}  \&
  {Zhang}}{{Wang} et~al.}{2022}]{wyn+2022}
{Wang} W.-Y.,  {Yang} Y.-P.,  {Niu} C.-H.,  {Xu} R.,   {Zhang} B.,  2022,
  \mn@doi [\apj] {10.3847/1538-4357/ac4097}, \href
  {https://ui.adsabs.harvard.edu/abs/2022ApJ...927..105W} {927, 105}

\bibitem[\protect\citeauthoryear{{Weisberg} et~al.,}{{Weisberg}
  et~al.}{1999}]{wcl+1999}
{Weisberg} J.~M.,  et~al., 1999, \mn@doi [\apjs] {10.1086/313189}, \href
  {https://ui.adsabs.harvard.edu/abs/1999ApJS..121..171W} {121, 171}

\bibitem[\protect\citeauthoryear{van Straten \& Bailes}{van Straten \&
  Bailes}{2011}]{dspsr}
van Straten W.,  Bailes M.,  2011, \mn@doi [pasa] {10.1071/as10021}, 28, 1

\bibitem[\protect\citeauthoryear{{van Straten}, {Demorest}  \& {Oslowski}}{{van
  Straten} et~al.}{2012}]{vdo+2012}
{van Straten} W.,  {Demorest} P.,   {Oslowski} S.,  2012, Astronomical Research
  and Technology, \href {https://ui.adsabs.harvard.edu/abs/2012AR&T....9..237V}
  {9, 237}

\makeatother
\end{thebibliography}





\appendix
\section{The polarization profiles of individual subpulses.}
\label{sec:SP}
The polarization data of all individual subpulses are available online here (link to data file). 

\begin{table*}
  \caption[]{Measured parameters for 955 single-subpulses of PSR B1916+14 observed by the FAST on May 28th, 2020 and March 8th, 2023. Those single-subpulses have a strong peak intensity larger than 20${\rm \sigma_I}$. Here we present the 10 single-subpulses and the entire table will be published online.}
    \label{tabB1}
\begin{tabular}{crrrrrrr}
    \hline \multicolumn{1}{c}{No.}             & \multicolumn{1}{c}{ $\phi_p$}         & \multicolumn{1}{c}{${\rm W_{sp}}$}      &  \multicolumn{1}{c}{${\rm I}$}  
       &  \multicolumn{1}{c}{${\rm L}$}  &  \multicolumn{1}{c}{${\rm V}$}  &  \multicolumn{1}{c}{${\rm |V|}$}  &  \multicolumn{1}{c}{${\Delta PA}$}\\ 
       &  \multicolumn{1}{c}{$(^\circ)$}       &   \multicolumn{1}{c}{(ms)}            &         \multicolumn{1}{c}{(Jy)}        &         \multicolumn{1}{c}{(Jy)}            &   \multicolumn{1}{c}{(Jy)}            &       \multicolumn{1}{c}{(Jy)}
       &  \multicolumn{1}{c}{$(^\circ)$}\\ 
          \multicolumn{1}{c}{(1)}              &     \multicolumn{1}{c}{(2)}           &        \multicolumn{1}{c}{(3)}          &      \multicolumn{1}{c}{(4)}       
       &  \multicolumn{1}{c}{(5)}              &      \multicolumn{1}{c}{(6)}          &         \multicolumn{1}{c}{(7)}         &         \multicolumn{1}{c}{(8)}\\ 
    \hline
 \multicolumn{8}{c}{2020-05-28}\\
\hline
No.001-1  &  -1.8  &  0.14  &  0.74 $\pm$ 0.02 &  0.64 $\pm$ 0.01 &  0.10 $\pm$ 0.02 &  0.10 &  -1.2 $\pm$2.2\\
No.002-1  &  -0.7  &  0.22  &  0.95 $\pm$ 0.03 &  0.77 $\pm$ 0.02 &  0.21 $\pm$ 0.03 &  0.21 &  6.7 $\pm$1.4\\
No.002-2  &  -0.0  &  0.27  &  1.45 $\pm$ 0.03 &  0.22 $\pm$ 0.02 &  0.35 $\pm$ 0.03 &  0.35 &  -4.3 $\pm$2.6\\
No.003-1  &  -1.6  &  0.24  &  0.93 $\pm$ 0.02 &  0.82 $\pm$ 0.02 &  0.00 $\pm$ 0.02 &  0.00 &  -3.5 $\pm$1.3\\
No.005-1  &  0.6  &  0.28  &  4.78 $\pm$ 0.03 &  4.39 $\pm$ 0.02 &  1.03 $\pm$ 0.03 &  1.03 &  13.6 $\pm$1.2\\
No.006-1  &  0.3  &  0.21  &  2.99 $\pm$ 0.03 &  1.91 $\pm$ 0.02 &  0.96 $\pm$ 0.03 &  0.96 &  -19.6 $\pm$1.4\\
No.006-2  &  1.3  &  0.17  &  3.25 $\pm$ 0.02 &  2.93 $\pm$ 0.02 &  1.11 $\pm$ 0.02 &  1.11 &  -0.2 $\pm$0.8\\
No.008-1  &  -3.9  &  0.19  &  0.92 $\pm$ 0.02 &  0.83 $\pm$ 0.01 &  0.06 $\pm$ 0.02 &  0.06 &  1.9 $\pm$1.3\\
No.008-2  &  1.0  &  0.36  &  3.33 $\pm$ 0.03 &  1.80 $\pm$ 0.02 &  0.68 $\pm$ 0.03 &  0.68 &  -22.3 $\pm$1.6\\
No.009-1  &  -3.4  &  0.17  &  1.26 $\pm$ 0.02 &  1.14 $\pm$ 0.02 &  0.06 $\pm$ 0.02 &  0.06 &  2.2 $\pm$0.9\\
\hline
\end{tabular}
\end{table*}

\section{The Method to recognize the single subpulses.}
\label{sec:SSP}

\begin{figure*}
\centering
\includegraphics[width=0.99\textwidth]{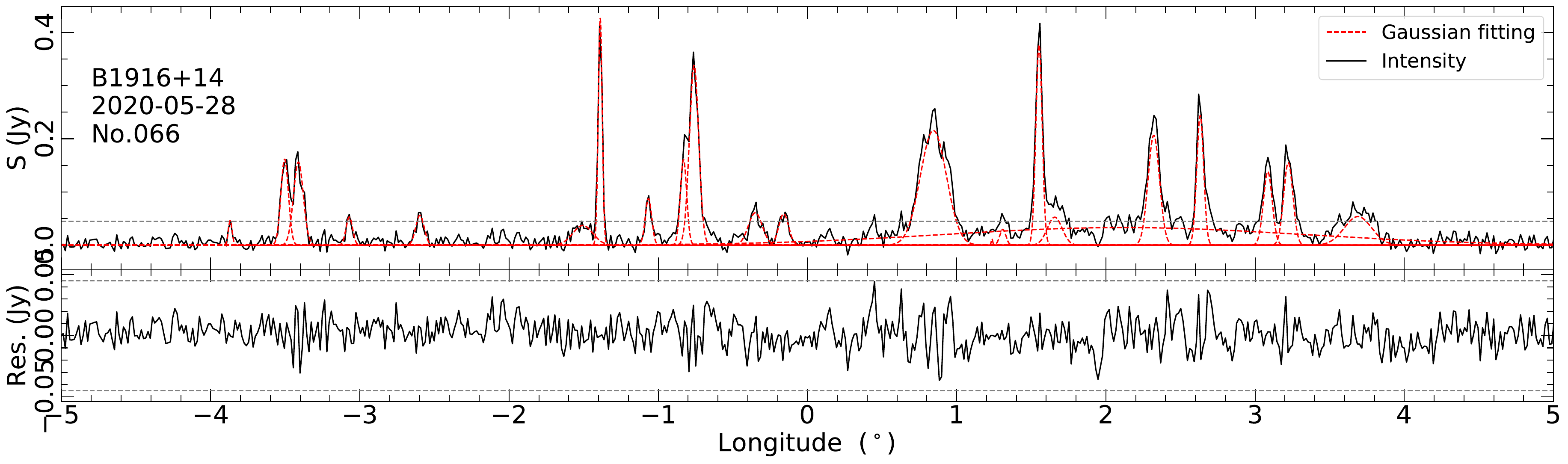}
\caption{The multi-Gaussian fittings to the individual pulse of pulsar period No.66 on 2020-05-28, as an example. The top panel shows the total intensity (solid line) with its 22 Gaussians (dashed lines), and the bottom panel exhibits the fitting residual. The dotted lines represent 5$\sigma_{I}$.}
\label{fit}
\end{figure*}

\begin{figure}
\centering
\includegraphics[width=0.45\textwidth]{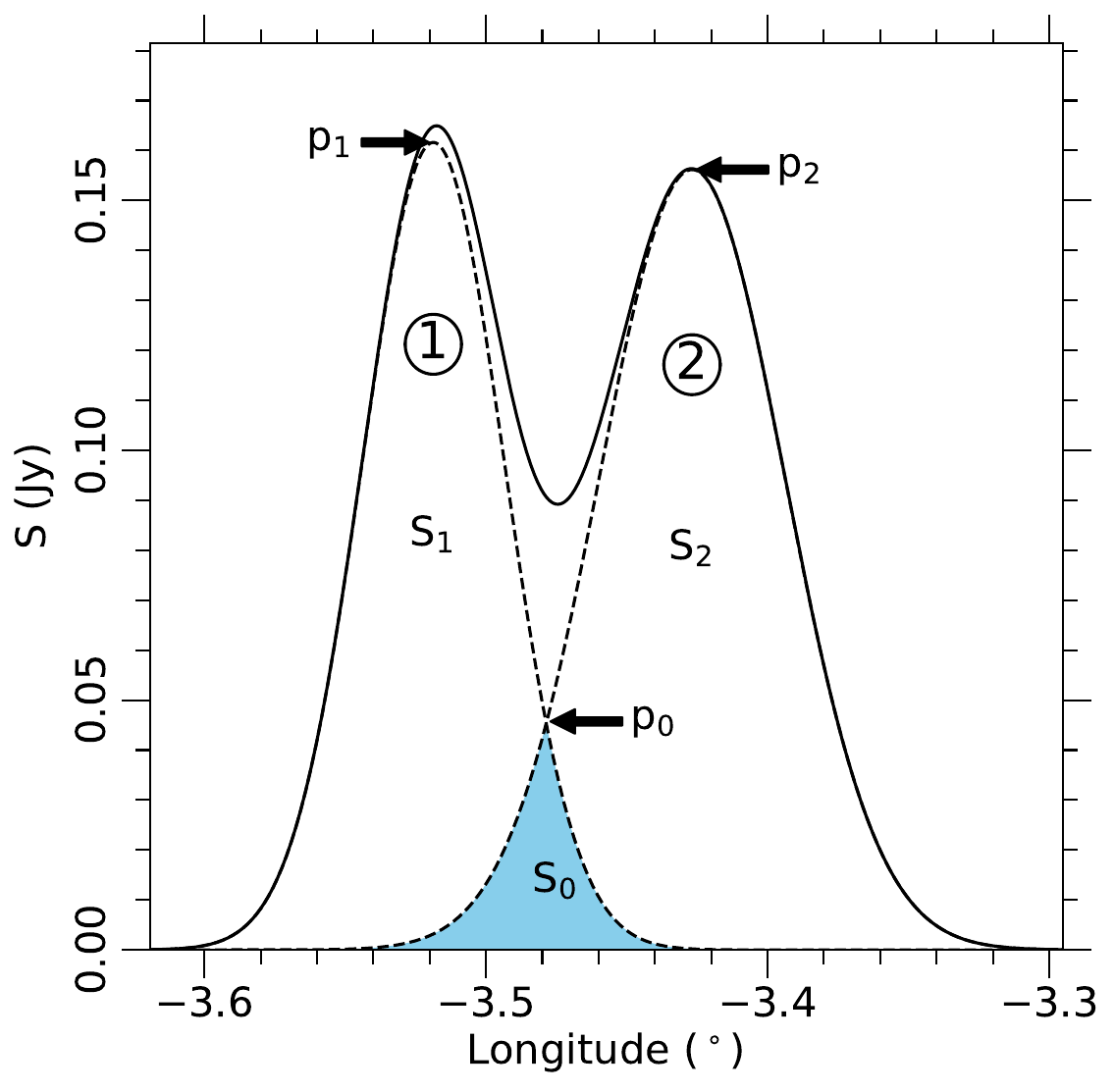}
\caption{Multi-Gaussian fittings to two slightly overlapped subpules in pulsar period No.66 on 2020-05-28. The p$_1$ and S$_1$ are the peak flux and the area of the 1st Gaussian shape respectively. So do for the 2nd one. The superposition part (the shadow area) between the two Gaussian shapes (dashed line) can be characterized by their cross point peak, p$_0$, and the superposition area, S$_0$. }
\label{cross}
\end{figure}

\begin{figure}
\centering
\includegraphics[width=0.45\textwidth]{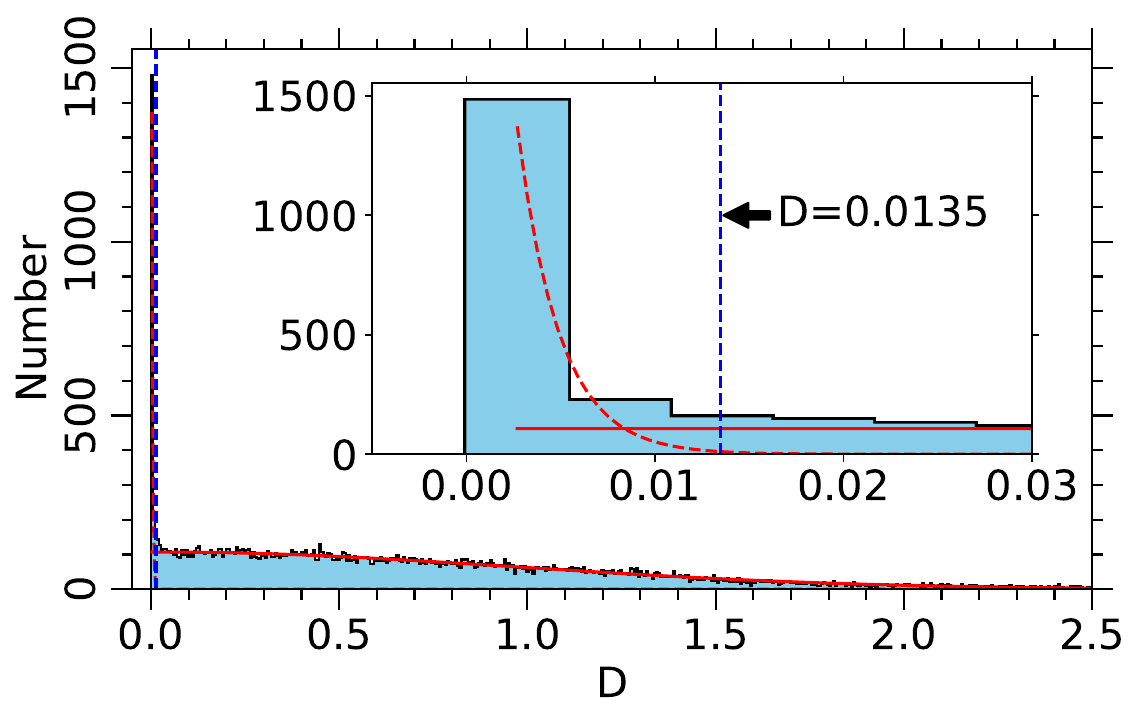}
\caption{The histogram of D for data of PSR B1916+14. It can be fitted well by a normal function (solid line) and an exponential function (dashed line). The dashed blue line limits the 99\% area of the exponential distribution where D=0.0135.}
\label{SSP}
\end{figure}

For PSR B1916+14, many individual pulses consist of two or more overlapping subpulses. We can fit them with multi-Gaussians until their residuals are smaller than 5$\sigma_{\rm I}$, as shown in Figure~\ref{fit} for an example of the pulse period No.66 on 2020-05-28. 

We get parameters for 23321 Gaussian components in all 1032 periods are observed by the FAST in two sessions, as listed in Table online. 

Figure~\ref{cross} shows an overlapped individual subpulse in pulse period No.66 on 2020-05-28, which is fitted two Gaussians (the dashed lines). Their superposition part (the blue area) can be characterized by the cross-point at p$_0$. For the 1st Gaussian shape, it is better to get a small p$_0$ and S$_0$ and hence a small D=${\rm \sqrt{ (p_0/p_1)^2+(S_0/S_1)^2}}$ to avoid any confusion of overlapped subpulses for better statistics. So does the 2nd Gaussian shape. 

We get the histogram of D for 23321 subpulses in 1032 periods as shown in Figure~\ref{SSP}. In this distribution, most of the subpulses follow a normal function (dotted line). Still, some of them go with an exponential function (dashed line) which represents those well-separated subpulses, or single-subpulses. We take D=0.0135 where the area of the exponential distribution reaches 99\% as a threshold to discriminate single-subpulses. Furthermore, to warrant high confidence for the polarization parameter determination, 955 single-subpulses whose peak values > 20$\sigma_{\rm I}$ are picked up in total. Their morphological and polarization parameters are measured as listed in Table~\ref{tabB1} (see Sec.\ref{sec:MPs} and \ref{sec:pol} for the details).


\bsp	
\label{lastpage}
\end{document}